\documentclass[aps, prd, reprint, nofootinbib, longbibliography, letterpaper, superscriptaddress, preprintnumbers]{revtex4-1}
\usepackage{graphics,graphicx,epsfig}
\usepackage{amsmath, amssymb, color}
\usepackage[dvipsnames]{xcolor}
\usepackage[colorlinks=true, linkcolor=blue, citecolor=magenta, hyperfootnotes=false]{hyperref}
\usepackage{url}
\usepackage{float, placeins}
\usepackage{mwe}
\usepackage[caption=false]{subfig}
\usepackage{morefloats}
\usepackage{empheq}
\usepackage{bm}
\usepackage{booktabs} 
\usepackage{multirow}
\usepackage{makecell}
\usepackage[normalem]{ulem}

\definecolor{dkgreen}{RGB}{0,110,0} 

\usepackage{ulem}

\def\calA{{\cal A}} 
\def\calB{{\cal B}}

\def\calD{{\cal D}}
\def\calE{{\cal E}}

\def\calI{{\cal I}}

\def\calO{{\cal O}}

\def\calP{{\cal P}}
\def\calS{{\cal S}}
\def\calT{{\cal T}}
\def\calN{{\cal N}}

\def\Uens{U^{\rm ens}}

\def\ens{{\rm ens}}
\def\Pcos{{P_{\rm cos}}}

\def\ntot{{n_{\rm tot}}}
\def\nprop{{n_{\rm prop}}}
\def\obs{{\rm obs}}
\def\Ureal{U_{\rm real}}
\def\pfirst{{p^{(1p)}}}

\def\={\hbox{\hskip 1pt$=$\hskip 1pt}} 
\def\expect#1{\left\langle #1\right\rangle}  

\usepackage{setspace}

\makeatletter

  \def\thesubsection{\thesection-\@Alph\c@subsection}

  \def\p@subsection{}

  \def\@sect@ltx#1#2#3#4#5#6[#7]#8{%
    \@ifnum{#2>\c@secnumdepth}{%
      \def\H@svsec{\phantomsection}%
      \let\@svsec\@empty
    }{%
      \H@refstepcounter{#1}%
      \def\H@svsec{%
       \phantomsection
      }%
      \protected@edef\@svsec{{#1}}%
      \@ifundefined{@#1cntformat}{%
       \prepdef\@svsec\@seccntformat
      }{%
       \expandafter\prepdef
       \expandafter\@svsec
                   \csname @#1cntformat\endcsname
      }%
    }%
    \@tempskipa #5\relax
    \@ifdim{\@tempskipa>\z@}{%
      \begingroup
        \interlinepenalty \@M
        #6{%
         \@ifundefined{@hangfrom@#1}{\@hang@from}{\csname @hangfrom@#1\endcsname}%
         {\hskip#3\relax\H@svsec}{\@svsec}{#8}%
        }%
        \@@par
      \endgroup
      \@ifundefined{#1mark}{\@gobble}{\csname #1mark\endcsname}{#7}%
      \addcontentsline{toc}{#1}{%
        \@ifnum{#2>\c@secnumdepth}{%
         \protect\numberline{}%
        }{%
         \protect\numberline{\csname thex#1\endcsname}%
        }%
        #8}%
    }{%
      \def\@svsechd{%
        #6{%
         \@ifundefined{@runin@to@#1}{\@runin@to}{\csname @runin@to@#1\endcsname}%
         {\hskip#3\relax\H@svsec}{\@svsec}{#8}%
        }%
        \@ifundefined{#1mark}{\@gobble}{\csname #1mark\endcsname}{#7}%
        \addcontentsline{toc}{#1}{%
          \@ifnum{#2>\c@secnumdepth}{%
           \protect\numberline{}%
          }{%
           \protect\numberline{\csname thex#1\endcsname}%
          }%
          #8}%
      }%
    }%
    \@xsect{#5}%
}

  \def\thexsubsection{\@Alph\c@subsection}

\makeatother

\begin{document}

\preprint{MIT-CTP-LI/5999}

\author{Feraz Azhar}
\email[Email address: {fazhar@nd.edu}]{}
\affiliation{Department of Philosophy, University of Notre Dame, Notre Dame, IN 46556, USA}

\author{Alan H. Guth}
\email[Email address: {guth@ctp.mit.edu}]{}
\affiliation{Department of Physics, Laboratory for Nuclear
Science, and Center for Theoretical Physics -- A Leinweber
Institute, Massachusetts Institute of Technology, Cambridge, MA
02139}

\author{Mohammad Hossein Namjoo}
\email[Email address: {mh.namjoo@ipm.ir}]{}
\affiliation{School of Astronomy, Institute for Research in Fundamental Sciences (IPM), Tehran,
Iran}


\title{Probabilistic inference in very large universes}

\date{February 2, 2026}

\begin{abstract}

Our current favored cosmological theories allow for the striking
and controversial possibility that the observable universe is
just a small part of a much larger universe in which parameters
that describe the effective, low-energy laws of physics vary from
one region to another. The controversy is largely driven by the
fact that such a ``very large universe'' is mostly
observationally inaccessible to us, so the issue arises of how we
can reasonably assess a theory that describes such a universe. In
this paper we propose a Bayesian method for theory assessment
based on theory-generated probability distributions for our
observations. We focus on the principles that define this method,
leaving aside concerns about how, in practice, one would carry
out the required calculations. (One important issue that we set
aside is the measure problem.) 

We argue that cosmological theories can be tested by the standard
method of Bayesian updating, but we need to use theoretical
predictions for ``first-person'' probabilities---that is,
probabilities that \emph{we} should use for \emph{our}
observations, taking into account all relevant selection effects.
These selection effects can vary from one observer to another,
and can vary with time, so, in principle, first-person
probabilities are defined for each \emph{observer-instant}---an
observer at a specific instant of time.  Calculations of
first-person probabilities should take into account everything
that the observer believes about herself and her surroundings,
which we refer to as her \emph{subjective state}. If the universe
is very large, a theory might predict that there are many
observer-instants in the same subjective state; we argue that
first-person probabilities should be calculated using a {\it
Principle of Self-Locating Indifference} (PSLI), the assumption
that any real observer should make predictions for her future as
if she were chosen randomly and uniformly from the theoretically
predicted observer-instants that share her subjective state. We
believe the PSLI is intuitively very reasonable, but we also
argue that, if the theory is correct, the use of this principle
maximizes the expected fraction of observers who will make
correct predictions. 

A further complication is that cosmological theories are not
expected to fully predict the detailed properties of the
universe, but rather will predict a set of possible universes,
each with a probability.  Different possible universes will
generically have different numbers of observers.  We argue that,
in the calculation of first-person probabilities, the probability
for each possible universe should be weighted by the number of
observer-instants in the specified subjective state that it
contains. 

These issues have been controversial in the literature, so we
also provide: a rebuttal to the claim that principles like the
PSLI involve a ``selection fallacy''; a rebuttal to what we dub
the {\it Principle of Required Certainty}; an argument rejecting
theories that predict a preponderance of Boltzmann brains; a
rebuttal to a parable about humans and Jovians used by Hartle and
Srednicki to argue that assumptions of typicality can lead to
absurd consequences; and finally, a discussion about how the use
of ``old evidence'' can be fit into a Bayesian mold.

\end{abstract} 

\maketitle

\def\refname{{\hskip -22.27782pt}References}
\def\acknowledgmentsname{{\hskip -22.27782pt}Acknowledgments}

\tableofcontents

\section{Introduction}
\label{SEC:Introduction}

A striking possibility suggested by the theories that underlie
our current understanding of the observable universe---such as
cosmic inflation~\citep{guth_81, linde_82,
albrecht+steinhardt_82, linde_83b}---is that the observable
universe is just one ``domain'' in a vast (mostly unobservable)
multi-domain structure (henceforth, a ``very large universe'',
also known as a ``multiverse'')~\citep{vilenkin_83, linde_86a,
linde_86b}. In such theories, the parameters that appear in the
effective, low-energy laws may vary from one domain to the next.
A central concern that arises in this context is: how do we test
theories of a very large universe?

There are of course many practical problems that arise in trying
to test theories of a very large universe, but there are also a
number of questions of principle that have remained controversial
in the literature.  In this paper we will focus on addressing
some of these questions of principle.  The first step, in our
view, is to specify the idealized methods of evaluating theories
of this type---i.e., methods that would be adopted if there were
no constraints on our ability to calculate the consequences of
the theory.  Then the development of practical methods should be
based on these idealized methods.

We assume that the theories in which we are interested are
sufficiently complete so that they can be used, in principle, to
calculate a probability distribution over the possible states of
the (very large) universe.  A crucial feature of the theories
that we have in mind is that they predict probabilities for the
state of the universe; they do not directly predict probabilities
for what any particular observer will measure.  In the context of
the theory, observers are simply objects that can arise as the
universe evolves.  Following Srednicki and Hartle
\cite{srednicki+hartle_10}, we will refer to these
theory-generated probabilities for the state of the universe as
``third-person'' probabilities.  If we were outside the universe,
and could see the entire universe, then we would be able to test
theories by comparing these third-person probabilities with the
universe or universes that we observe.  The problem, however, is
that we are not outside the universe, but instead are among the
objects that are produced within the universe.  Furthermore, if
the theory predicts (with some probability) the existence of
observers who could be us, then it may very well predict the
existence of many such observers, and the theory will provide no
information about which copy actually corresponds to us. So, to
be able to test such a theory, we need to be able to use the
third-person probabilities predicted by the theory to infer
``first-person'' probabilities---probabilities for {\em our}
observations.\footnote{Third-person and first-person perspectives
have a rich history (see, for examples from various disciplines,
Refs.~\cite{perry_79, lewis_79, bostrom_02, hawking+hertog_06,
hartle+hertog_15}). The first authors to explicitly highlight
these terms in the context of probabilistic reasoning in a very
large universe, we believe, were Srednicki and Hartle in
Ref.~\cite{srednicki+hartle_10}. In earlier papers, starting in,
e.g.,~\citet{hawking+hertog_06}, such third-person and
first-person probabilities have also been referred to as
``bottom-up'' and ``top-down'' probabilities, respectively.}

One problem that arises in defining first- and third-person
probabilities is known as the ``measure problem.'' It involves
issues with defining a probability measure over sample spaces in
which events of interest occur an infinite number of times. We
will not attempt to address the measure problem in this
paper. (See Refs.~\cite{aguirre_07, aguirre+al_07, guth_07, freivogel_11, smeenk_14} for a discussion of this problem.)  Thus our paper will be directly relevant to universes that are
very large, but nonetheless have finite spacetime volumes, where
there generally are no measure problems.  We expect, however,
that our claims will also be applicable for infinite universes
once a solution to the measure problem has been adopted.

We adopt a standard Bayesian framework for testing theories, in
which the probabilities assigned to theories are updated
according to the results of experiments.  To calculate the
updated probabilities, however, we need to know the probabilities
that the theory being tested predicts for the possible outcomes
of the experiment.  Since these are probabilities for experiments
to be performed by us, they are unambiguously first-person
probabilities. 

To understand the importance of the first-person perspective in
predicting the outcomes of physics experiments, we can consider a
model of a very large universe in the context of string theory,
in which different regions in this very large universe are built
upon different types of vacuum.  We can imagine that many vacua
are consistent with our current data, but each of these vacua
might lead to different probabilities for the outcomes of future
experiments.  So the predictions of any theory for experiments
that we will perform in the future (which are first-person
probabilities) will depend on what the theory tells us about the
probabilities for the different types of vacua in which we might
reside.  The bottom line is that we cannot test theories unless
we have a methodology for understanding the predictions of
theories for first-person probabilities, taking into account all
relevant selection effects.%
\footnote{According to McMullin
\citep{McMullin1993}, ``A selection effect is a feature found in
what is observed, due to the circumstances of the observation or
to the means used, rather than being intrinsic to the object
observed.'' Note that the ``features'' mentioned in this
definition do not have to be qualitatively new characteristics,
but might simply be shifts in statistical distributions.}

A key problem that arises in determining first-person
probabilities is the ``typicality problem,'' which concerns the
issue of whether, given a solution to the measure problem, we
should assume that we (and thus our future observations) are
\emph{typical}~\cite{hartle+srednicki_07, page_08,
srednicki+hartle_10, azhar_14, azhar_15, azhar_16}. By typical,
we mean that probabilities for our future observations can be
obtained by treating ourselves as if we were randomly and
uniformly chosen from some specified set of observers. There are
two distinct positions that have been argued for in the
literature. The first position, dubbed the ``principle of
mediocrity'' by~\citet{vilenkin_95}, asserts that we should
reason as though we are indeed typical~\citep{gott_93,
vilenkin_95, bostrom_02, page_07, garriga+vilenkin_08}---this can
be understood as a version of the ``principle of
indifference''~\cite{keynes_21}. The second position asserts that
typicality is just one assumption among a ``spectrum'' of
possible assumptions enforcing various degrees of (a)typicality,
with the principle of indifference corresponding to a single such
assumption~\citep{hartle+srednicki_07, srednicki+hartle_10,
azhar_14, azhar_15, azhar_16, azhar+butterfield_18}. 

In this paper we argue for the first position---that we should
proceed under the assumption of typicality. Typicality has
traditionally been the assumption underlying probabilistic
reasoning, and it has so far been very successful.  When theories
make only probabilistic predictions, there is no method of
analysis that can guarantee that correct conclusions will result. 
However, we will argue that the assumption of typicality will
always lead to the largest expectation value for the fraction of
observers who reach correct conclusions.  We will also argue
that, without an assumption of typicality, it would be
essentially impossible to exclude {\em any} theory on the basis
of observation.

Once we decide that we are ``typical,'' we face another important
question: what is the ``reference class'' within which we can
assume that we are typical.  This is also known as the
``conditionalization problem''.  That is, probabilities for our
future measurements are often assigned under the assumption that
we are typical of some reference class of observers, where the
probabilities are based on the fractions of the reference class
that are predicted to obtain any particular experimental result. 
One approach to choosing an appropriate reference class is is to
employ anthropic constraints, namely, conditions that are viewed
as being necessary for the existence of any observers (in the
style of Carter's Weak Anthropic Principle~\cite{carter_74}), but
such constraints are difficult to make precise.  Another option
is to conditionalize on all previous data we have accumulated,
but it remains necessary to define exactly what is meant by
``data''. (See
Refs.~%
\cite{garriga+vilenkin_08, hartle+hertog_13,
aguirre+tegmark_05,azhar+butterfield_18, azhar+linnemann_25} for
further discussion.)

In this paper, we will adopt a particular idealized method of
conditionalization, based on the goal of avoiding arbitrary
choices whenever possible.  So, to avoid having to decide if
Neanderthals are enough like us to be counted in the same
reference class, we advocate that the reference class should be
taken as narrowly as possible.  (In Sec.~\ref{APP:OldEv},
however, we will have further comments about the possible
relevance of larger reference classes.) We would expect that such
a reference class would lead to the most accurate possible
predictions, for it makes use of all the information available to
us---thereby precluding the charge of biasing the predictions by
ignoring relevant facts, or even possibly cherry-picking the data
to find a desired result. This choice means roughly that we
should conditionalize our predictions on observers who are {\it
exactly} like us.

But we are not exactly like each other, nor are we even exactly
like ourselves at different times.  So we are led to the
proposition that predictions should in principle be made
separately for each observer, and for each point in
time.\footnote{In Ref.~\cite{garriga+vilenkin_08}, p.~3, Garriga
and Vilenkin give a toy example of a situation in which a
prediction depends strongly on the observer.  A universe is
populated by civilizations that are each one of two sizes, large
or small, where the number of large and small civilizations are
known to be equal.  For some reason, nobody has been able to
count, or even estimate, the population of her civilization.  It
is also known that each civilization, large or small, has exactly
one astronomer.  In this situation, the astronomers would assign
a probability of 1/2 for being in a large or small civilization,
since half of the astronomers live in each type of civilization. 
But each ordinary citizen would consider herself more likely to
be in a large civilization, since most ordinary citizens live in
large civilizations.  While these two probabilities differ, each
correctly describes the predictions that can be made for the
observers for whom the probabilities were calculated.}

The reader might at this point be concerned that if we draw the
reference class so narrowly, then maybe the only member of the
class would be the observer for whom the predictions are sought,
and nothing could be learned.  However, we must keep in mind that
we are discussing theory-generated probabilities, and not
probabilities based on the frequency of actual observed events.
As long as the theory predicts a nonzero probability for an
observer meeting some particular description $\calD$, it is in
principle possible to infer {\em conditional} probabilities
$p(\calE|{\calD})=p(\calE, \calD)/p(\calD)$ for the events
$\calE$ that an observer meeting this description will
experience. ($p(\calE, \calD)$ denotes the probability of $\calE$
{\it and} $\calD$.)

We use the symbol ${\cal O}$ to denote an ``observer,'' a term
that we use to describe any physical entity that has the capacity
to record and utilize information (similar to an IGUS---an {\it
information gathering and utilizing system}---discussed
by~\citet{gell-mann+hartle_89}; see also
Ref.~\cite{srednicki+hartle_10}).\footnote{We are assuming that
all properties of observers (or at least all relevant properties)
can be described by the laws of physics.\label{FN:Observers}}
Much of our discussion will revolve around {\it
observer-instants} $\,\calI$, where $\,\calI$ denotes the doublet
$(\calO_,t)$, specifying the observer $\calO$ at a {\it
particular} time $t$.\footnote{This concept is similar to
Bostrom's ``observer-moment'' \cite{bostrom_02}.  An
``observer-moment,'' however, refers to a brief time segment of
an observer, while we use ``observer-instant'' to refer to a
particular time $t$. While nothing {\it happens} at an instant of
time, we focus on the state of the observer, which is
well-defined at an instant $t$.} We will use the notation
$\calO(\calI)$ to denote the observer associated with
observer-instant $\,\calI$, and $t(\calI)$ to denote the time
associated with $\,\calI$.

To make predictions for an observer-instant $\,\calI$, we want to
conditionalize on observer-instants exactly like it. But there is
a further complication, in that nobody knows, or can know, how to
identify observer-instants {\it exactly} like some particular
observer-instant $\,\calI$.  An observer $\calO$ can examine her
own memories and thought processes, but she cannot determine her
precise physical state.  She cannot really be sure of her own
thought processes, either, since she can examine her thought
processes only by relying on them.  If she were insane, and maybe
even if she were not, she may have a radically deluded image of
herself. One could even imagine that the observer $\calO$ is
actually a disembodied brain kept alive in a vat of nutrients,
with all sensory nerves connected to a computer that presents the
brain with a totally artificial world. 

Since the observer $\calO$ cannot know her precise physical
state, we use the term \emph{subjective state} to describe what
she actually believes about herself and her surroundings.  We
therefore define

\begin{quote}
{\bf Subjective state%
.} The subjective state of an observer instant $\,\calI \equiv
(\calO,t)$ is the set of all properties that the observer
$\,\calO$ believes, at time $t$, about herself and her
surroundings.  These beliefs may or may not be true, and include
her memories, her self-conception of her thought processes, and
anything else that she is aware of about herself or her
surroundings.  We denote this set of properties by
$\calS(\calI)$, or simply $\calS$ when $\,\calI$ need not be
specified.
\end{quote}

\noindent We will 
assume that the best predictions that an observer $\calO$ can
make will be those that are conditionalized on her subjective
state at the current time, i.e., the subjective state $\calS$ of
her current observer-instant $\,\calI$.  This point of view
agrees with Garriga and Vilenkin
\citep[Sec.~II]{garriga+vilenkin_08}, who advocated that
``ideally'' one should use a reference class of ``observers with
identical information content,'' which should be ``characterized
by the full information content at her disposal.''

We will later discuss the use of an \emph{ensemble} of universes
constructed from a theory, which is a list of a very large number
(approaching infinity) of fully detailed universe descriptions,
generated with the probability distribution implied by the
theory.  By ``fully detailed,'' we mean that the description
completely specifies not only the cosmological features of the
universe, but also the complete histories of everything that
occurs in the entire spacetime of the universe. To determine
first-person probabilities for a given real observer, it will be
necessary to identify observer-instants in the ensemble that are
in the same {\em subjective state} as the real observer's current
observer-instant.  Although we do not believe that such an
ensemble of fully detailed universe descriptions can actually be
constructed, we believe that thinking about such an ensemble is a
good way to describe how probabilities should in principle be
extracted from theories. In standard classical statistical
mechanics, after all, we discuss ensembles of $10^{23}$ particles
without worrying that the description of one element of this
ensemble would, as of 2025, exceed the total storage capacity of
all the computers on planet Earth.

We also introduce a concept which we refer to as LIKEness,
defined in terms of the concept of {\em subjective state}:

\begin{quote}
{\bf LIKE.} Two observer-instants $\,\calI$ and $\,\calI'$ are
LIKE each other if they share the same {\em subjective state}
$\calS$.  (``LIKE'', when used in this way, will always be
capitalized.)
\end{quote}

Since $\calS$ includes everything that each observer believes
about herself or her surroundings, if an observer-instant
$\,\calI'$ is LIKE an observer-instant $\,\calI$, then the
observer $\calO(\calI)$ cannot, using only information available
at time $t(\calI)$, distinguish her current observer-instant
$\,\calI$ from $\,\calI'$, no matter how much she learns about
$\,\calI'$. If $\calO(\calI)$ is a real observer and $\,\calI'$
is an observer-instant in the theoretically-generated ensemble,
then $\calO(\calI)$ might know more about $\,\calI'$ than its
subjective state.  For example, $\calO(\calI)$ would be able to
tell if $\,\calI'$ is a brain in a vat. This additional
information, however, cannot affect her judgment about whether
$\,\calI'$ is LIKE her, since she does not know anything about
her own observer-instant beyond her subjective state; i.e., she
does not know whether she herself is a brain in a
vat.\footnote{We understand that some philosophers argue that we
{\em can} tell that we are {\em not} brains in a vat
\cite{Moore1959,
Reid1983}; or---at least---we can tell that the external world,
as we perceive it, exists
\cite{Nozick1981}. 
We, however, will be interested in theories of very large
universes that predict the existence of Boltzmann brains.  In the
context of these theories, it seems clearly rational to take
seriously the idea that a conscious being might have no reliable
information about the external world.}

Finally, we point out that the LIKEness of two observer-instants,
according to our definitions, is an objective property---a third
person with full access to the memories, thought processes, and
other aspects of the subjective state of two observer-instants
$\,\calI$ and $\,\calI'$ can in principle decide unambiguously if
$\,\calI'$ is LIKE $\,\calI$. 

To summarize, we propose that an observer making predictions for
her current observer-instant $\,\calI$ should conditionalize on
observer-instants $\,\calI'$ that are LIKE $\,\calI$.  Other
observers can also make predictions for $\,\calI$, and should
also make use of everything they believe about $\,\calI$.  But
they might know more or less about $\,\calI$ than $\calO(\calI)$
does, so they would in general make different predictions.  The
methods that we will describe apply to any observer making
predictions for any other, but we will focus on predictions made
by an observer for herself.  Our main goal is to describe how to
test theories in very large universes, and an experimenter will
do this by comparing her experimental results with the
predictions implied by the theory
for \emph{her} experiments.%
\footnote{\label{FN:semiclassical}Our description here
has been entirely classical.  We are assuming that a classical
description of observers can effectively capture their relevant
features, though we recognize that (i) a complete physical
description would require a specification of their quantum state
and (ii) we do not have a complete understanding of how the
classical picture emerges from the quantum description.  Phrased
another way, we accept the experience of our everyday lives, in
which the world that we observe, despite its underlying
quantum-mechanical nature, is described well by a classical
picture. For example, the evolution of very large universes that
arise through the nucleation of ``bubbles'' can mostly be
described by classical general relativity, despite the fact that
the nucleation event is intrinsically quantum mechanical and must
be described probabilistically. On a much smaller scale, in an
experiment involving the radioactive decay of an atom, the local
``click'' of a Geiger counter can be described in terms of
classical probabilities, despite this event being intrinsically
quantum mechanical.}

If the universe is sufficiently large, then it would be expected
to contain many observer-instants LIKE any particular $\,\calI$.
But even if the universe is small, it is still very reasonable to
condition on observer-instants LIKE $\,\calI$.  As discussed
above, if we are talking about theory-generated probabilities,
then as long as the theory predicts a nonzero probability for
such an observer-instant, the conditional probabilities for any
observation, given that the observer-instant is LIKE $\,\calI$,
are well-defined.

Another important feature of the theories that we have in mind is
that the number of observer-instants that are LIKE any particular
$\,\calI$ can vary from one possible outcome (viz., one possible
very large universe in the ensemble) to another. The question
arises: how should we weight each such outcome in our
computations of the first-person probabilities for the future of
an observer-instant $\,\calI$? One suggestion asserts that we
should weight each outcome in proportion to its (third-person)
probability of occurrence, while restricting to outcomes that
contain at least one instance of an observer-instant that is LIKE
$\,\calI$. 

As will be discussed more completely in Sec.~\ref{SEC:PSLI}, we
argue that predictions for a given observer-instant $\,\calI$
should be calculated according to what we dub the {\it Principle
of Self-Locating Indifference} (PSLI)---i.e., they should be
calculated as if the observer-instant were chosen randomly and
uniformly from the set of all observer-instants
in the ensemble that are in the same subjective state.%
\footnote{%
This statement has to be worded carefully.  Note that our
statement of the PSLI is \emph{not} equivalent to
\begin{quote}
Statement A: Each of the observer instants $\,\calI_i$ in the
ensemble that are LIKE $\,\calI$ is equally likely to have a
future that matches that of $\,\calI$. 
\end{quote}
Contrary to Statement A, the PSLI implies that a particular
observer instant $\,\calI_i$ is more likely to have a future that
matches that of $\,\calI$ if there are many different
$\,\calI_i$'s that share the same future.  Thus, we contend that
Statement A is false.  Similarly, the PSLI is \emph{not}
equivalent to
\begin{quote}
Statement B: The observer instant $\,\calI$ is equally likely to
have a future that matches that of any of the observer instants
$\,\calI_i$ in the ensemble that are LIKE $\,\calI$.
\end{quote} 
Statement B is equivalent to Statement A, since the statement
that $\,\calI_i$ has a future that matches that of $\,\calI$ is
equivalent to the statement that $\,\calI$ has a future that
matches that of $\,\calI_i$. Thus, Statement B is also false. 
Our statement of the PSLI is not phrased in terms of the equality
of probabilities, but can be summarized by the somewhat vague
statement that all of the observer-instants $\calI_i$ that are in
the same subjective state as $\calI$ contribute equally to the
predictions for $\calI$.} This principle leads to weighting each
model universe in proportion to the number of observer-instants
LIKE $\,\calI$ that
it contains.%
\footnote{ Note that this
weighting by the number of observers does not commit us to what
is known as the ``self-indication assumption''
(SIA).~\citet[p.~66]{bostrom_02} defines the SIA as: ``Given the
fact that you exist, you should (other things equal) favor
hypotheses according to which many observers exist over
hypotheses on which few observers exist''. Such an assumption
increases prior probabilities for hypotheses in which there are a
larger number of observers and leads to the patent absurdity
known as the ``presumptuous philosopher'' problem
(see~\citet[p.~124]{bostrom_02})---the problem that any
hypothesis can claim a high prior probability simply by
presumptuously appending an assumption that it gives rise to a
colossal number of observers.  We endorse weighting by the number
of observers in the calculation of the first-person probabilities
predicted by a given theory, but this does not lead to this
absurdity. }

The PSLI can be implemented in operational terms by what we call
the
{\it Method of Many Repetitions}%
.  In this context we imagine some proposition pertaining to an
observer-instant, such as the proposition that the card that was
just drawn by the observer $\calO(\calI)$ was a jack of spades,
or the proposition that $\calO(\calI)$ is about to be hit by
lightning.  The method stipulates that the first-person
probability for any such proposition is equal to the fraction of
observer-instants in the same subjective state in the ensemble
for which the proposition is true.

The plan for this paper is as follows. In
Sec.~\ref{SEC:BayesianTheory}, to be explicit about the general
probabilistic setting in which we will be working, we describe
the standard Bayesian framework for theory assessment. In
Sec.~\ref{SEC:PIMMR2}, we develop and motivate the {\it Principle
of Self-Locating Indifference} for computing first-person
probabilities for observables, as well as the {\it Method of Many
Repetitions}. In Sec.~\ref{SEC:Other} we provide further thoughts
on probabilistic inference in very large universes, including: a
rebuttal to the claim that principles like our PSLI involve a
``selection fallacy''; a rebuttal to what we dub the {\it
Principle of Required Certainty}; an argument rejecting theories
that predict a preponderance of Boltzmann brains; a rebuttal to a
parable about humans and Jovians used by Hartle and Srednicki
\cite{hartle+srednicki_07} to argue that assumptions of
typicality can lead to absurd consequences; and finally, a
discussion about how the use of ``old evidence'' can be fit into
a Bayesian mold. Concluding remarks follow in
Sec.~\ref{SEC:Conclusion}. In Appendices~\ref{APP:PSLIknown}
and~\ref{APP:simp_assump}, we provide further technical details
that motivate the {\it Principle of Self-Locating Indifference}.
And, in Appendix~\ref{APP:BoxMod}, we provide an explicit
comparison of first-person probabilities calculated via (a) an
application of the {\it Principle of Self-Locating Indifference}
and (b) methods used by Hartle and collaborators in
Refs.~\cite{hartle+srednicki_07}, \cite{srednicki+hartle_10},
\cite{hartle+hertog_15}, and \cite{hartle+hertog_16}%
---highlighting differences in the results obtained.

To summarize our overall position, we contend that despite the
novelty inherent in theories that describe a very large universe,
these novelties can be treated pragmatically by the same Bayesian
methods that are commonly used in empirical physics.  We assume
the {\it Principle of Self-Locating Indifference}, but we regard
this assumption as a natural extension of the typicality
assumptions that have always been used in statistical data
analysis.  This is an optimistic stance (!), considering how
little we know about the possibility of a multiverse. 

\section{Bayesian theory assessment}\label{SEC:BayesianTheory}

We will use a Bayesian framework for theory assessment, which we
describe in this section%
---mainly to establish notation. When Bayes' theorem is used to
update the probabilities that we assign to our theories, we will
need to use first-person probabilities, because they are the
relevant probabilities for our measurements.  First-person
probabilities were described briefly in the Introduction, and a
precise proposal for calculating them will be described in the
following section.

In general terms, Bayes' theorem can be derived from the
definition of conditional probability. For two propositions, $A$
and $B$, provided that $P(B)\neq 0$, the probability of $A$ given
$B$ is defined as $P(A|B) \equiv P(A,B)/P(B)$, where $P(A,B)$
is the probability of $A$ {\it and} $B$.  Similarly, provided
that $P(A)\neq 0$, $P(B|A) \equiv P(B,A)/P(A)$.  Since $P(A,B) =
P(B,A)$,
\begin{equation}\label{EQN:OrigBAYES}
P(A|B) = \frac{P(B|A)P(A)}{P(B)},
\end{equation}
provided that $P(B)$ and $P(A)$ are both nonzero.%
\footnote{Note that the probabilities in Bayes' theorem are
sometimes explicitly written as being conditional on some
background knowledge---comprising information that can affect
$P(A|B)$~\cite{jaynes_03}. For simplicity we suppress such
notation throughout---but assume that probabilities can indeed
depend on such background knowledge, provided that all relevant
probabilities are conditioned on the same background knowledge.}

One important application of Eq.~(\ref{EQN:OrigBAYES}) describes
how new data, $D_{\rm new}$, impacts the probabilities that we
assign to theories $\mathcal{T}_{i}$.  If we set, in
Eq.~(\ref{EQN:OrigBAYES}), $A = \mathcal{T}_{i}$ and $B = D_{\rm
new}$, we obtain
\begin{equation}\label{EQN:BayesK}
P(\mathcal{T}_i|D_{\rm new})=\frac{P(D_{\rm new}|\mathcal{T}_i)P(\mathcal{T}_i)}{P(D_{\rm new})},
\end{equation}
provided that $P(D_{\rm new})$ is nonzero. 

The quantity on the left-hand side of Eq.~(\ref{EQN:BayesK}),
namely $P(\mathcal{T}_i|D_{\rm new})$, is the probability that we
assign to $\mathcal{T}_{i}$ after $D_{\rm new}$ is obtained, and
is known as the \emph{posterior probability} of $\mathcal{T}_i$.
The probability that we assign to $\mathcal{T}_i$ prior to our
learning $D_{\rm new}$, $P(\mathcal{T}_i)$, is known as the
\emph{prior probability} (or just the `prior') of
$\mathcal{T}_i$.  Thus, Bayes' theorem provides a means to update
one's confidence in a theory based on new data.  The probability
$P(D_{\rm new}|\mathcal{T}_i)$, the theory-generated probability
for $D_{\rm new}$ given the theory $\mathcal{T}_i$, is often
called the `likelihood'. In general, it may be very difficult to
calculate this quantity, and this difficulty can serve as a
limitation on our ability to evaluate a theory.  In this paper,
however, we will consider the idealization in which all such
quantities are assumed to be calculable.

Finally, if we also assume that $\mathcal{T}_i$ belongs to an
exhaustive and mutually exclusive set of theories
$\{\mathcal{T}_{1}, \mathcal{T}_{2}, \dots, \mathcal{T}_{N}\}$
(see Refs.~\cite{jeffreys_83, earman_92, bernardo+smith_94,
jaynes_03, howson+urbach_06, lee_12}), one can use the law of
total probability to write Bayes' theorem as
\begin{equation}\label{EQN:BayesKlaw}
P(\mathcal{T}_{i}|D_{\rm new})=\frac{P(D_{\rm new}|\mathcal{T}_{i})P(\mathcal{T}_{i})}
{\displaystyle \sum_{j=1}^{N}P(D_{\rm new}|\mathcal{T}_{j})P(\mathcal{T}_{j})}\ .
\end{equation}

\section{Calculating first-person probabilities in very large
universes}\label{SEC:PIMMR2}

We begin by assuming that we are exploring the predictions of a
fully detailed cosmological theory. An (ideal) complete
cosmological theory provides a set of possible universes
$\{U_\alpha\}$, where each $U_\alpha$ is a {\it full description}
of a possible spacetime {\em and} everything that happens in it. 
Each $U_\alpha$ describes an entire universe, even if the
universe has disconnected components.  (Some theories may predict
universes with disconnected components, and others may not.  We
are allowing for both possibilities.) Each $U_\alpha$ is fully
detailed, describing not only the cosmological features of the
universe, but also the complete histories of every being and
every object in the universe, including every experiment that is
ever done, every word that is ever uttered, every thought that is
ever thunk, and every star, planet, meteorite, and speck of dust
that ever exists. As described in Footnote
\ref{FN:semiclassical}, we are assuming that, for these purposes, a classical description suffices.  
We assume that $U_\alpha$ describes the universe modulo
coordinate redefinitions, so two descriptions that are related by
a coordinate redefinition correspond to the same $U_\alpha$. (In
this paper, as we described in the Introduction, we will consider
explicitly only the case in which the universe has a finite
spacetime volume, although we expect our conclusions to also
apply to infinite universes, once a measure is adopted.)

In this language, the physical world that we observe---which we
denote by $\Ureal$---is theorized to be the realization of one of
the $\{U_\alpha\}$, meaning that $\Ureal$ is fully described by a
particular model universe $U_\alpha$. The model universe
$U_\alpha$ that is realized will be denoted by $U_*$.

The theory does not tell us which of the $U_\alpha$ will be
realized, and we do not have any expectation that we will ever
gather enough information to learn which $U_\alpha$ is realized. 
But the theory in principle allows us to calculate a probability
function, $\Pcos(U_\alpha)$, which describes the probability that
the real universe, $\Ureal$, is described by $U_\alpha$
(equivalently, that $U_\alpha=U_{*}$). (The subscript ``cos''
stands for ``cosmological''.) The theoretical framework of
interest relies on the assumption that it is meaningful to assign
probabilities to such unverifiable outcomes, just as classical
statistical mechanics relies on the assumption that it is
meaningful to assign probabilities to $\calO(10^{23})$
phase-space variables, even though we know that we could never
measure all of these variables. Thus, while we can never directly
test any prediction for the probability of a particular model
universe
     $U_\alpha$, as specified by $\Pcos(U_\alpha)$, knowledge of
     $\Pcos(U_\alpha)$ for all the $U_\alpha$'s can be used to
extract probabilistic predictions for observables that we {\it
will} be able to test.\footnote{For simplicity we will treat the
label $\alpha$ as a discrete variable, but it would be
straightforward to generalize our arguments to the case in which
the possible universes are labeled by a set of variables, any of
which may be discrete or continuous.}

It may at first seem inadequate to assign probabilities only to
fully described universes $U_\alpha$, while not explicitly
assigning probabilities to events within such universes. To
understand why only these full-universe probabilities are needed,
imagine that the label $\alpha$ can be written explicitly as a
huge set of real-valued parameters.  Then the specification of
the probabilities $\Pcos(U_\alpha)$ would amount to a
specification of the joint probability distribution for these
parameters. The probability distributions for any subset of
parameters can be found by marginalizing the full joint
probability distribution over the other parameters.

\subsection{Probabilities, third person and first person}\label{SEC:ProbTF}

The probabilities $\Pcos(U_\alpha)$ describe the probability
distribution for the universe from a \emph{third person} point of
view---the point of view of an external observer (who is outside
of the universe) and can observe the entire universe.  However,
the theory says nothing whatever about who {\it I} am, so we have
to make further assumptions to make predictions about what {\it
I} (or any other particular observer) should expect to
experience.  The probabilities that describe what any particular
observer should expect to experience are called first-person
probabilities. 

In more detail, even if we assume that the theory is correct and
that we know which $U_\alpha$ describes the real universe $U_{\rm
real}$, it is quite possible in a very large universe that
$U_{\rm real}$ will contain multiple observer-instants that are
LIKE any given observer-instant $\,\calI_*$ for which predictions
are sought.  That is, it is possible that $\Ureal$ contains
multiple observer-instants $\,\calI_i$ that are in the same
subjective state as $\,\calI_*$, $\calS(\calI_*)$, so that the
observer $\calO(\calI_*)$ (who is \emph{inside} the universe)
cannot at time $t(\calI_*)$ distinguish any of them from herself.
Furthermore, since we don't expect that anyone will actually know
which $U_\alpha$ describes the real universe $U_{\rm real}$,
there is the further complication that different possible
universes $U_\alpha$ might each have different numbers of
observer-instants $\,\calI_i$ which are in the subjective state
$\calS(\calI_*)$.  In this situation, the following question
arises:
\begin{quote}
How should the observer $\calO(\calI_*)$ compute probabilities
for what she will experience?
\end{quote}
That is, how should one calculate the \emph{first-person}
probabilities for what the observer $\calO(\calI_*)$ will
experience?

\subsection{\textit{Principle of Self-Locating Indifference}}
\label{SEC:PSLI}

To describe the answer that we propose to this question, we first
define some terms that will be useful.  Recall that we are
attempting to describe the consequences of a fully detailed
cosmological theory, which provides a set of possible physical
states (i.e., the set of possible universes $\{U_\alpha\}$) and a
probability distribution $\Pcos(U_\alpha)$ over these physical
states. 

Following discussions that are common in statistical mechanics
textbooks, we will imagine using the theory to generate a large
list of model universes $\Uens_1, \Uens_2, \ldots,
\Uens_N$---called an \emph{ensemble}.  Each entry $\Uens_\lambda$
on the list is one of the possible universes $\{U_\alpha\}$,
chosen according to the probability distribution
$\Pcos(U_\alpha)$.%
\footnote{Gibbs \cite{gibbs_1902} originally defined the
notion of an ensemble in exactly this way, as a ``great number of
independent systems,'' each in a definite state described by a
point in phase space. The density of these points in phase space
was used instead of any notion of probability.}

Note that the ensemble is not a set, since a given model universe
$U_\alpha$ can appear in the ensemble multiple times, while a
given element can appear in a set only once. Note also that the
ensemble is not uniquely determined by the theory, since each
entry is chosen randomly. We will be interested, however,
exclusively in the statistical properties of the ensemble in the
limit $N \rightarrow \infty$, which {\it are} uniquely defined by
the theory.  To be clear, we coin the term {\it asymptotic
ensemble} to refer to an ensemble that is described as a list of
$N$ entries, but for which the only properties that one is
allowed to extract are those that arise from the statistical
distribution in the limit $N \rightarrow
\infty$.  Thus, each theory determines a unique asymptotic
ensemble.

The ensemble is not essential to our methods, but it simplifies
the discussion by encoding the probability distribution
$\Pcos(U_\alpha)$ into the multiplicities of universes $U_\alpha$
in the ensemble.  Thus, choosing an entry randomly and uniformly
from the asymptotic ensemble is equivalent to choosing a model
universe $U_\alpha$ according to the probability distribution
$\Pcos(U_\alpha)$.

If we know the asymptotic ensemble implied by a theory, and if
somebody asks us the probability that $U_\alpha$ is the universe
that is realized, we can immediately answer $\Pcos(U_\alpha)$,
without any further thought.  But to compare the predictions of
the theory with observations, we need to be able to answer a much
wider class of questions.  Since each $U_\alpha$ is a complete
description of the universe, we should be able to answer
questions about probabilities for anything that can happen within
a universe.

We are interested in constructing a method to calculate
probabilities for arbitrary questions that an observer might ask
about her future, or even her past, or for events in regions that
are space-like separated from her current observer-instant.  In
the classical approximation to reality that we are describing,
the entire spacetime around the observer has a well-defined
description, whether the observer has direct access to these
regions of spacetime or not. 

Our goal is to define a method to calculate the probability for
any proposition of interest pertaining to an observer at a
particular instant in time.  We want to be able to calculate, for
example, the probability that a specific event happens in the
future or the past of some observer-instant, or perhaps the
probability that a specific set of such events happen in the
future or the past of some observer-instant. It could also be, as
another example, the probability that a distant vacuum decay
event occurs (one that was initially spacelike separated from the
observer), which sets off an expanding bubble wall that will
collide with the observer 5 years into the future.  The
probabilities of interest can be time-specific (e.g., the
probability that the observer will have a car accident 22 days in
the future), or not time-specific (e.g., the probability that the
observer will have a car accident sometime in her future). 

We can now state the principle that we advocate to calculate such
first-person probabilities. We will first state the principle,
discussing motivation and justification below.

\begin{quote}
{\bf Principle of Self-Locating Indifference (PSLI)\null.} An
observer should calculate first-person probabilities for any
proposition pertaining to her current observer-instant as if the
observer-instant were chosen randomly and uniformly from the set
of all observer-instants in the asymptotic ensemble (generated
from the theory) that are in the same subjective state.
\end{quote}

We can illustrate this principle via a simple example. Consider a
scenario in which our fully detailed cosmological theory allows
two possible universes $U_1$ and $U_2$, with probabilities
$P_{\rm cos}(U_1)=1/3$ and $P_{\rm cos}(U_2)=2/3$.  Assume that
we are interested in predictions for an observer-instant whose
current subjective state is $\calS$. Let there be one
observer-instant in $U_1$ and two observer-instants in $U_2$ that
are in the subjective state $\calS$.  Suppose that the (lone)
observer of interest in $U_1$ suffers a myocardial infarction
within the next 24 hours; while one of the observers of interest
in $U_2$ suffers a myocardial infarction in the next 24 hours. We
may then ask:
\begin{quote}
If an observer's current subjective state is $\calS$, what is the
probability that the observer will have a myocardial infarction
within the next 24 hours?
\end{quote}
To answer this question, we construct the theory-generated
ensemble---a list of a large number $N$ of universe descriptions,
as discussed above.  If the list is sufficiently large, we expect
$U_1$ to appear on the list about $N/3$ times and $U_2$ to appear
about $2N/3$ times.%
\footnote{Our style here follows Gibbs' example, discussing an
ensemble with a finite number $N$ of elements, while relying on
statements that are only true in the limit $N \rightarrow
\infty$.  In a footnote on p.~5 of Ref.~\cite{gibbs_1902}, Gibbs
wrote ``In strictness, a finite number of systems cannot be
distributed continuously in phase [phase space]. But by
increasing indefinitely the number of systems, we may approximate
to a continuous law of distribution, such as is here described.''
Gibbs goes on to say that such language, ``although wanting in
precision of expression,'' will be used ``to avoid tedious
circumlocution \dots when the sense in which it is to be taken
appears sufficiently clear.''}
There will be about $N/3$ observer-instants in subjective state
$\calS$ in universes of type $U_1$ and about $4N/3$ such
observer-instants in universes of type $U_2$.  Since about 3/5 of
these $5N/3$ observer-instants suffer a myocardial infarction
within the next 24 hours, an observer-instant chosen randomly and
uniformly from this set will have a probability of 60\% of having
a myocardial infarction within the next 24 hours.  Thus,
according to the PSLI, the real observer of interest should
predict that she has a 60\% probability of having a myocardial
infarction within 24 hours.

This example may be generalized to a statement which we dub the
{\it Method of Many Repetitions}.  The {\it Method of Many
Repetitions} is logically equivalent to the PSLI, but its wording
more directly describes how first-person probabilities are to be
operationally calculated.

\begin{quote}
{\bf Method of Many Repetitions (MMR).\null } The first-person
probability for any proposition pertaining to a particular
observer-instant is the fraction of the observer-instants in the
same subjective state, in the theory-generated asymptotic
ensemble, that belong to observers for whom the proposition
holds. 
\end{quote}

The MMR is an operational procedure to calculate the consequences
of the PSLI.  In particular, it can be used to derive an explicit
equation for the first person probability $\pfirst(\calP |
\calS)$ that any proposition ${\calP}$ will hold for an
observer-instant in subjective state $\calS$.  Once the theory of
interest has been used to calculate the set of possible universes
$\{U_\alpha\}$ and their probabilities $\Pcos(U_\alpha)$, we can
let $\ntot(U_\alpha, \calS)$ denote the total number of
observer-instants in subjective state $\calS$ in the model
universe $U_\alpha$, and we can let $n_{\rm prop}(U_\alpha,
\calS, \calP)$ denote the number of observer-instants in
subjective state $\calS$ in the model universe $U_\alpha$ for
which the proposition $\calP$ is true. If $N$ is the total number
of model universes in the ensemble, then the expected total
number of observer instants in subjective state $S$ in the
ensemble is given by
\begin{align}
\calN_{\rm tot}(\calS) =  N \sum_\beta P_{\rm cos}(U_\beta)\,
     \ntot(U_\beta, \calS)\ .
\end{align}
Of these observer-instants, the expected number for which the
proposition $\calP$ is true is given by
\begin{align}
\calN_{\rm prop}(\calS, \calP) =  N \sum_\beta P_{\rm cos}(U_\beta)\,
     \nprop(U_\beta, \calS, \calP)\ .
\end{align}
The MMR tells us that the first-person probability that
proposition $\calP$ is true for an observer-instant in subjective
state $\calS$ is given by the ratio of these two quantities,
\begin{align}
\pfirst(\calP|\calS) &= \frac{\calN_{\rm prop}(\calS,\calP)}
     {\calN_{\rm tot}(\calS)}\\
     &= \frac{\displaystyle \sum_\beta P_{\rm cos}(U_\beta)\,
     \nprop(U_\beta, \calS, \calP)}
     {\displaystyle \sum_\beta P_{\rm
     cos}(U_\beta)\,\ntot(U_\beta, \calS)} \ .
     \label{EQN:P1pP}
\end{align}
This prescription for calculating first-person probabilities
based on the PSLI is one of the main results of this paper. 
Since it fully describes how to calculate the first-person
probability for an arbitrary proposition $\calP$, it is fully
equivalent to the definition of the PSLI (written out in words as
above).

Finally, the above formalism allows us to address the question of
the first-person probability that $\calO(\calI)$, for a given
observer-instant $\,\calI$ in subjective state $\calS$, lives in
a universe described by a specific $U_\alpha$. If
$\mathcal{P}_\alpha$ is the proposition that the specified
observer-instant $\,\calI$ lives in a universe described by
$U_\alpha$, then $\nprop(U_\beta, \calS, \mathcal{P}_\alpha) =
\ntot(U_\alpha,\calS) \delta_{\alpha \beta}$,
so Eq.~(\ref{EQN:P1pP}) implies that
\begin{align}
     \pfirst(U_{\alpha}|\calS) = \frac{
     P_{\rm cos}(U_\alpha)\,
     \ntot(U_\alpha, \calS)}
     {\displaystyle \sum_\beta P_{\rm
     cos}(U_\beta)\,\ntot(U_\beta, \calS)} \ .
     \label{EQN:P1p_Ualpha}
\end{align}
Importantly, Eq.~\eqref{EQN:P1p_Ualpha} states that the
probability that I live in a particular universe $U_\alpha$ is
not only proportional to the probability $\Pcos(U_\alpha)$ that
the universe is realized (according to the theory), but it is
also proportional to the number $\ntot(U_\alpha, \calS)$ of
observer-instants LIKE mine in that universe. The latter
factor---that is, the weight associated with the number of
observer-instants---is an `anthropic' effect, resulting from the
PSLI. 

To justify the procedure that we are proposing to define
first-person probabilities, in the following section we will
discuss the underlying motivation for principles of indifference. 
For readers inclined to accept indifference as a reasonable
assumption, these arguments might be sufficient.  However, in
Appendices~\ref{APP:PSLIknown} and \ref{APP:simp_assump} we
attempt to give more detailed arguments based on more primitive
assumptions.

In Appendix~\ref{APP:PSLIknown} we make the simplifying
assumption that we know which model universe $U_* \in
\{U_\alpha\}$ is actually realized.  Even if $U_*$ is known,
however, the calculation of first-person probabilities for some
real observer-instant $\,\calI$ is complicated by the possibility
that the model universe $U_*$ might contain any number of
observer-instants that are LIKE $\,\calI$, and are therefore
indistinguishable from $\,\calI$ (to the observer $\calO(\calI)$
at the time $t(\calI)$). Thus, we need to specify the probability
that $\,\calI$ might be described by any particular one of them. 
Since the goal is to define a procedure for extracting
first-person predictions from theories in very large universes,
we will use the success of such predictions as our criterion for
comparing different possible procedures. In particular, in
Appendix~\ref{APP:PSLIknown} we will show that the use of the
PSLI to define first-person probabilities maximizes the expected
fraction of observers who will deduce correct inferences from
their observations. 

In Appendix~\ref{APP:simp_assump} we will drop the simplifying
assumption that $U_*$ is known, and we will give an argument,
based on what we consider very reasonable assumptions, that the
first-person probability for any particular model universe
$U_\alpha$ should be weighted by the number of observer-instants
in the model universe that are LIKE $\,\calI$.  These arguments
in Appendix~\ref{APP:simp_assump} will allow us to rederive
Eq.~(\ref{EQN:P1pP}), a key result of this paper which is
equivalent to the PSLI.

\subsection{Motivating indifference}
\label{SEC:MotivatingIndifference}

Our key principle, the PSLI, mandates that observers ought to
calculate probabilities for outcomes of interest as if their
current observer-instant were chosen randomly and uniformly from
those observer-instants in the same subjective state in the
asymptotic ensemble generated from the theory.  This method of
making predictions for a real observer implements the assumption
that her observer-instant is typical of the observer-instants in
the ensemble that are in the same subjective state. So we must
address the question: when is it appropriate to assume that
predictions should be made as if the case of interest is typical
of some reference class? Some have argued that we should consider
ourselves typical of any reference class for which there is no
evidence to the contrary. Others have argued that there is no
evidence that we are typical of any group, and that we should
therefore allow for the possibility that we are simply not
typical.

The question of our typicality is an example of the more general
question of deciding under what circumstances the probabilities
of alternative outcomes should be taken to be equal.  The attempt
to answer this more general question has led to a variety of
points of view, including various formulations of the {\it
principle of indifference.} The principle of indifference has its
origins in work by~\citet{bernoulli_06} and~\citet{laplace_02}.
The term `principle of indifference' was coined by
\citet{keynes_21}, who also proposed a more precise description
of the qualifications that are needed before a set of outcomes
can be assigned equal probabilities.  For further discussion, see
Ref.~\cite{russell_1992}. 

For the present discussion we only need a principle of
indifference that is adapted to the self-location problem
discussed in the previous section. Here we adopt the PSLI\null. 

Note that principles of indifference adapted to the self-location
problem in cosmology are apparently not (yet) universally
accepted among cosmologists. They have been challenged in at
least two ways.

First, a possible position is that if an observer does not know
which observer she is (in some appropriate reference class), then
she should just accept the fact that she does not know the
probabilities.  We agree that it is always possible to adopt the
position that we do not know the probabilities. Similarly, if
somebody asks us what is the capital of Kazakhstan, we can
honestly reply that we do not know.  ``We do not know'' is a
possible answer to most any question, but it does not exclude the
possibility of other answers that are also correct and possibly
more informative.  While the observer may not know which of the
observers she corresponds to, the PSLI is a way of quantifying
her uncertainty, based on the available information.  We can
sometimes be forced to make such judgments.  If a tornado is
approaching, we may have to decide in which direction to run,
which would presumably involve some estimate of the probabilities
of different paths that the tornado might follow.  We would make
the best estimate we could, based on available information, and
there would be no value in maintaining the position that we do
not know the probabilities. 

Second, principles of indifference adapted to the self-location
problem have been challenged, most emphatically in
Ref.~\cite{srednicki+hartle_10}, wherein it is claimed that
typicality is ``no more motivated by observation than
atypicality.''

We agree that there are no observational tests of our typicality
in the universe, but we believe that there is a solid argument
for typicality, based on counting.

The counting argument that we have in mind can be illustrated in
a standard textbook probability problem involving coin flipping. 
If a million people each flip a fair coin 20 times, there are
better than even odds that at least one will flip 20 heads in a
row.  But neither probability theory, nor anything else that we
know about the laws of physics tells us anything about which of
the million people is likely to be the one who flips 20 heads in
a row.  In evaluating the implications of the probability
distribution, however, we always assume that {\it we} are very
unlikely to be the one-in-a-million coin flipper who flips 20
heads in a row.  More generally, if we consider 1.5 million
coin-flippers, probability theory tells us that most likely there
will be at least one flipper who flips exactly $n$ heads, for
every $n$ from 0 to 20. If we did not make the assumption that we
are typical, then all we would know is that any $n$ is possible,
and we would lose all the probabilistic predictions of coin
flipping.

The counting argument, in the context of betting, was presented
very clearly by Vilenkin in Ref.~\cite{vilenkin_11}.  Vilenkin
imagines a conference in which all attendees have been given
hats, 80\% white and 20\% black, but nobody has been able to see
the color of his or her own hat. To register for the conference,
everyone is required to bet $\pounds$100 on the color of her hat.
Everyone can look around and notice that the hats are about 80\%
white.  Thus, if the attendees all assume that they are typical
and therefore bet on white, 80\% will win their bets.  By
contrast, if the attendees all assume that they don't know the
probabilities, they might place a random bet and on average only
half would win.  Clearly, the assumption of typicality is the
better betting strategy.  While the 80\% vs.\ 50\% in Vilenkin's
example may not sound compelling, the advantage of assuming
typicality can be made more dramatic by considering repetitions
of this scenario.  If the same group attends 20 conferences with
the same protocol and assume that they are typical, 99.7\% will
have net winnings.  By contrast, if they bet randomly, again on
average only half would win.

In Appendix \ref{APP:PSLIknown}, as we summarized above, we will
show explicitly that Vilenkin's white hat/black hat parable
generalizes to the case under discussion here---the evaluation of
theories.  We will show that analyzing experimental data using
the PSLI (or equivalently the {\it Method of Many Repetitions)}
maximizes the expectation value of the fraction of experimenters
who will reach correct conclusions from their data. 
This is our take on the issue of typicality.

Note that as emphasized by Garriga and Vilenkin
\cite[p.~3]{garriga+vilenkin_08}, such assumptions of typicality
are routine in standard laboratory settings. Say, for example, an
experimenter measures the value of some physical quantity,
obtaining the result $m$.  Suppose further that we are interested
in comparing two theories, $\mathcal{T}_1$ and $\mathcal{T}_2$,
in light of this experimental result. The standard Bayesian
approach is to compare theory-generated probabilities for $m$
(viz., likelihoods), given a description of the experiment, where
this description is denoted by $E$. That is, one compares
$P(m|\mathcal{T}_1, E)$ with $P(m|\mathcal{T}_2, E)$. If
$P(m|\mathcal{T}_2, E) \ll P(m|\mathcal{T}_1, E)$ (that is, if
the likelihoods strongly favor $\mathcal{T}_1$), then the
standard conclusion would be that $\mathcal{T}_2$ is strongly
disfavored. We would not expect it to be claimed that
$\mathcal{T}_2$ is perfectly acceptable since we can freely
assume that we are among the extremely atypical observers who
obtain the result $m$, despite its low probability.  In other
words, in standard laboratory settings, assumptions of
atypicality are never introduced to avoid disfavoring a theory
with a very small likelihood.

In Ref.~\cite{srednicki+hartle_10}, Srednicki and Hartle propose
that it is not appropriate to imagine that we can test theories
directly, but rather we should always think of testing {\it
frameworks,} where a framework consists of a pair $(\calT, \xi)$,
where $\calT$ is a theory, and $\xi$ is a {\it xerographic
distribution}.  (They assume that the full description of the
actual universe is known, to focus on the question of our place
in the universe.) If our data $D_0$ is replicated at spacetime
locations $\{x_1, x_2, \dots , x_N\}$, then the xerographic
distribution is an assumed (i.e., not derived from a cosmological
theory) set of probabilities $\{\xi_1, \xi_2, \dots,\xi_N\}$ that
we are located at each of the spacetime locations $\{x_1, x_2,
\dots , x_N\}$, respectively.  We would argue that if one allows
xerographic distributions in which we are highly atypical, then
empirical science as we know it would become essentially
impossible.  To see this, consider a theory $\calT$ for which
$P(D_0|\calT)$, the probability of our data $D_0$ given the
theory $\calT$, is extremely small.  If $P(D_0|\calT)$ is
nonetheless large enough so that $D_0$ is likely to occur at
least once in the very large universe, we can always arrange for
$P(D_0|\calT, \xi)$ to be nearly one, by choosing a xerographic
distribution $\xi$ which implies that we are among the very rare
observers who observe $D_0$.  (This is analogous to assuming that
we are likely to be the 20-head coin flipper mentioned earlier in
this section.) In a very large (or possibly infinite) universe, a
huge range of theories are likely to be found acceptable by this
criterion.

While some may conclude that we are forced to accept this huge
range of theories, we do not see any justification for this
conclusion.  Recall, (i) the expectation value of the number of
observers who make correct deductions is maximized by using the
assumption of typicality, which then allows us to drastically
narrow the range of acceptable theories; (ii) such assumptions of
typicality are routine in standard laboratory settings.

\section{Other thoughts on probabilities in very large universes}
\label{SEC:Other}

In this section we discuss five issues related to the general
topic of probabilistic reasoning in the context of very large
universes.

\subsection{The \textit{Selection Fallacy} fallacy}
\label{SEC:Fallacy}

In Ref.~\cite{hartle+srednicki_07}, Hartle and Srednicki wrote
\begin{quote}
``To compute likelihoods as though we had been randomly selected
by some physical process, when there is no evidence for such a
process, commits what might be called the {\it selection
fallacy}. We are not a disembodied entity that was randomly
selected to have a particular physical description; instead, we
are the meaning of a transcription of `we' into the language of
physical theory.''
\end{quote}

Here we argue that this claim of a fallacy is in fact false.  To
assume that we are typical of some reference class does not
require us to assume that we are a disembodied entity that was
randomly selected by some process to match some particular member
of that reference class.  All that is necessary, we argue, is to
know that we are a member of the reference class, and to have no
reason to believe that we are any more likely to be one
particular member rather than another. 

To make our point, let us consider a thought experiment involving
the hypothetical cloning of humans.  Suppose a subject is put to
sleep, and while asleep five clones of the subject are made, and
for simplicity (recognizing that this is only a thought
experiment), we assume that the original subject is killed.  We
then have five clones, all identical, with nothing to distinguish
them other than their {\it creation number,} from 1 to 5, which
indicates the order of their creation.  They are placed into
identical rooms, numbered 1 to 5, according to their creation
number.  They are all awakened, but are not told their creation
number and cannot see the numbers on their rooms.

In addition, each clone is also assigned a randomized number,
also from 1 to 5.  There are 120 permutations of the numbers from
1 to 5, so the randomized numbers are assigned by flipping a fair
120-sided coin, with each outcome linked to a permutation.  The
permutation is then used to assign the randomized numbers.  For
example, if the randomly chosen permutation is (4,2,3,5,1), then
the clones in rooms 1 through 5 will be assigned the randomized
numbers 4, 2, 3, 5, and 1, respectively.

Now we consider the probabilities that each clone should assign
to their creation numbers and their randomized numbers.  Hartle
and Srednicki's {\it selection fallacy} argument would seem to
imply that the creation numbers cannot be treated as random,
since there was no random process involved in determining them. 
(They are closely analogous to the observer-instant numbers that
we use in Appendix
\ref{APP:simp_assump}.)  They are simply an indication of the
room in which the clone is sitting, so the creation number is
purely a question of self-location.

On the other hand, the randomized numbers were indeed chosen
randomly, so surely probability theory should apply here.  Each
clone should regard his randomized number as having a probability
of 1/5 to be any number from 1 to 5.

We argue, however, that there is no relevant difference between
these two cases (i.e., the two number assignments).  If someone
offers all the clones an opportunity to bet on their creation
number or randomized number with better than 5:1 odds, it would
pay on average for them to take the bet.  If they all bet on
``5'' for either their creation number or randomized number,
exactly one will win and four will lose, but the winnings of the
winner would exceed the total of the losses.  Most importantly,
there would be no difference between the two cases.

As is usual in probabilistic predictions, the predictions can be
improved significantly with repetition.  The cloning experiment
can be repeated by making 5 clones of each of the clones, and
then making 5 clones of each of those clones, etc.  After each
cloning the original subject is destroyed, but each of the 5
clones considers himself to be the continuation of the original
subject---each clone acquires all the memories of the original
subject, and interprets these memories as memories of his own
past.

As a concrete example, let's suppose that the experiment is
repeated 100 times.  At the end there will be $5^{100}$ clones,
which are a lot of mouths to feed, but since this is a thought
experiment we will not worry about this hypothetical hunger
emergency.  At each repetition, each of the clones is offered the
opportunity to place separate bets on his creation number and his
randomized number.  If he guesses wrong, he loses \$1.00; if he
guesses right, he wins \$10.00, so the odds are in his favor.  At
these odds, clones who guess right at least 10 of the 100 times
will come out ahead.  Suppose that on each repetition they all
bet on the same choice as each other, on both the creation number
and randomized number.  Since 1 in 5 will guess right on each
trial, the fraction of clones who guess right exactly $n$ times
out of $N$ trials is given by the binomial formula as
\begin{equation}
F(n) = \frac{N!}{n! (N-n)!} \left(\frac{1}{5}\right)^n \left(
     \frac{4}{5}\right)^{N-n} \ .
\end{equation}
Using this formula to sum the fractions from $n=10$ to $n=100$,
with $N = 100$, we find that under these assumptions 99.77\% of
the clones will come out ahead on their bets, for either the
creation number bets or the randomized number bets. Again, there
is no difference between the two cases.  If we discard the
assumption that all clones bet on the same choice as each other
in each repetition, and instead assume that they bet randomly,
then we conclude that the expectation value for the fraction of
clones that come out ahead is 99.77\%.  Again, all predictions
are the same for the two cases.

At this point the {\it selection fallacy} advocates might say
that of course there is no difference between self-location and
random choice if one only looks at global averages, but if I were
part of this experiment, only one of these clones would be {\it
me}.  For the randomized numbers, they would say, probability
theory convinces me that I am very unlikely to be part of the
0.23\% who lose on their bets, while for the creation numbers I
don't know this.  But if pressed to describe the physical
difference between these two cases, we are not aware of any
sensible answer that the {\it selection fallacy} advocate might
give. 

In fact, we will argue that the clones would have absolutely no
way to distinguish between their creation numbers and their
randomized numbers.  To be concrete, let us assume that after
each bet is concluded, each clone is told the correct value of
his creation number and randomized number.  At the end of the
experiment there are then $5^{100}$ clones, each of which has
recorded a sequence of 100 creation numbers, each from 1 to 5,
and a sequence of 100 randomized numbers, also from 1 to 5. 
Suppose we imagine a variant of the experiment in which the
clones are told that they will each be assigned a creation number
and a randomized number, as in the original version, but in this
case the numbers will be presented as X and Y\null.  The clones
are told only that the symbols will be used consistently---X will
always be the creation number, and Y the randomized number, or
maybe the other way around. In this case, we see no way that the
clones could determine which sequence was which---even if they
had the opportunity to pool all of the data from the $5^{100}$
clones.  We believe that in either case, we can rely on the
belief that if only 0.23\% of the group will lose money, and I
have no way of knowing where in the group I will fall, then I can
assume that it is very unlikely that I will lose money. Thus, for
either sequence, it pays for the clones to place their bets.

We would further argue that the physical description of the
system after 100 clonings allows no place for {\it me} to be
special.  Each possible creation number sequence occurs once and
only once, and the same can be said for the randomized number
sequences.  They were assigned by different methods, but their
description at the end of the experiments seems to be the same in
all relevant aspects.

The physics of the system knows nothing about who {\it I} am. 
Rather, the physics of the system includes $5^{100}$ different
``{\it I}$\,$''{}'s, where each ``{\it I}$\,$'' refers to the
thoughts happening in one of the $5^{100}$ heads of the clones. 
The physical situation associates one of the $5^{100}$ ``{\it
I}$\,$''{}'s with each possible creation number sequence, and
similarly one of the $5^{100}$ ``{\it I}$\,$''{}'s with each
possible randomized number sequence. Physics, as far as we know
it, is always formulated in terms of a description of possible
physical states, and a prescription for how such states evolve
with time.  Within that framework, we cannot see any way that any
one of the clones can be treated as special, not subject to the
same probability distribution as the rest. 

Finally, we point out that in practice there is no doubt that the
world accepts the use of probabilistic reasoning in situations
where there is no ``random process'' giving rise to the
uncertainty.  In Sec.~\ref{SEC:Jovians}, especially in Footnote
\ref{FN:medical}, we will emphasize the medical application. 
Here we highlight the use of Monte Carlo calculations, which have
become ubiquitous in physics and in many other fields.  The
``random numbers'' used in these calculations are actually
pseudorandom---they are generated by a purely deterministic
algorithm, but the algorithm is chosen to be sufficiently complex
so that there is no practical way to predict the next bit that
will be generated with better than even odds.  Thus, although
there is no random process, there is also no practical way to
conclude that the next bit is more likely to be a 0 or 1, even
after observing a large number of output digits and applying a
suite of pattern-searching algorithms. In our view this is all
that is needed, and all that we often have, to justify the use of
probabilistic reasoning based on indifference.

\subsection{When we evaluate theories, should we use only what we
know for certain?}\label{SEC:PRC}

When we perform a measurement obtaining data $D_0$, the only fact
about the universe that we know {\em for certain} (assuming that
we can trust our analysis) is that there exists at least one
instance of $D_0$ in the universe.  In
Ref.~\cite{hartle+srednicki_07}, Hartle and Srednicki argued that
in evaluating theories, we should use only this certain
information.  As described in Sec.~\ref{SEC:PIMMR2}, we
(strongly) disagree with this assertion, but for purposes of
discussion we give it a name---the {\it Principle of Required
Certainty}.

\begin{quote}
{\bf Principle of Required Certainty (PRC)%
.} When we perform a measurement obtaining some data $D_0$, only
the fact that we know for certain---the fact that at least one
instance of $D_0$ exists in the universe---should enter into our
evaluation of theories of the universe.
\end{quote}

\noindent By contrast, a more standard Bayesian analysis would
take into account the probabilistic inferences stemming from the
fact that we have done one experiment of this type, and found the
result $D_0$.

In the authors' own words (Ref.~\cite{hartle+srednicki_07},
p.~2),
\begin{quote}
All we know is that there exists at least one such region [region
with exactly our data] containing our data.
\end{quote}
In evaluating the implications of a measurement, these authors
base all of their reasoning {\it solely} on this point of
certainty. 

After stating that the existence of at least one instance of
$D_0$ is all we know for certain, the authors of
Ref.~\cite{hartle+srednicki_07} proceed to reason on the basis of
this fact as if it were standard in science to rely only on what
we know for certain.  In our view, however, this is an extremely
inaccurate description of how science proceeds. Scientific
reasoning has never been limited to what we know for certain, and
indeed science as we know it would come to a halt if such a
restriction were adopted. 

If we look at the experimental literature, we often see $1\sigma$
error bars with their approximately 68\% confidence level, and
sometimes $2\sigma$ error bars with their approximately 95\%
confidence level.  But we have {\em never} seen error bars of
certainty.  Indeed, if experimenters were required to use error
bars of certainty, experimental papers would essentially
disappear.  Similarly, we have seen papers that exclude certain
hypotheses at the 95\% confidence level, and sometimes the 99\%
confidence level.  But we have {\em never} seen a scientific
paper that discusses the exclusion of a hypothesis at the 100\%
confidence level.

To see how the requirement of certainty is at odds with
conventional Bayesian reasoning, we can consider the following
textbook problem, from the textbook by
A.~V.~Skorokhod~\cite{skorokhod_1989}:
\begin{quote}
There are two urns of which the first contains 2 white and 8
black balls and the second 8 white and 2 black balls. An urn is
selected at random and a ball is drawn from it. It is white. What
is the probability that the first urn was chosen?
\end{quote}
The standard answer, as given in the textbook, is that our
selection of a white ball favors the second urn, in which the
fraction of white balls is larger, so the probability of the
first urn is found to be only 1/5. On the other hand, if we
approached this question by first asking what we know {\it for
certain} about the urns, the answer would be that the draw of a
white ball tells us only that the urn contains at least one white
ball.  Since we already knew that both urns contain at least one
white ball, we have learned nothing.  The PRC would therefore
imply that the probability that we chose the first urn would be
unchanged from its prior, which was taken to be 1/2. Thus, the
PRC is in contradiction with conventional Bayesian reasoning. 
(Note that we are not saying that Hartle and Srednicki
\cite{hartle+srednicki_07} have incorrectly applied Bayes'
theorem.  Rather, we are saying that they departed from
conventional Bayesian reasoning by using only the fact that the
universe contains at least one instance of our data.)

To see how the PRC affects the analysis of an experiment,
consider an effort to measure the decay rate of a particle $X$. 
Suppose that we put $N$ of them in a bottle, and in time $\Delta
t$ we observe that $k$ of them decay. We assume the conventional
model of particle decays: each particle that has not already
decayed is assumed to have a probability $\lambda \, dt$ of
decaying in any time interval $dt$, independent of the
environment or prior history. Provided that $k \ll N$, we would
estimate the decay rate to be $\lambda = k/(N \, \Delta t)$, and
an estimate of the $1\sigma$ uncertainty would be $\Delta\lambda
\approx
\sqrt{k}/(N \, \Delta t)$. But suppose, instead, we asked what do
we learn {\it for certain} about the decay rate? Is it possible,
for example, that the actual decay rate is smaller by a factor of
a billion, or larger by a factor of a billion? Yes, although the
result that we found would have been very unlikely in that case.
But it is not impossible.  So the only thing that we learn {\it
for certain} is that the decay rate is nonzero.  This situation
is not changed if we imagine repeating the experiment many (but
only a finite number) of times.  If we pool all the data from all
the repetitions, we can ask the same questions, and they will
have the same answers.

In the standard Bayesian analysis, by contrast, we never learn
for certain what the decay rate is, but if the experiment
involves a high number of counts, it will strongly favor the
theory in which the decay rate is near $k/(N\,\Delta t)$. Physics
as we know it depends crucially on such probabilistic inferences. 
If only certainties were allowed in our scientific reasoning, we
would know essentially nothing.

One important consequence of the PRC is to dramatically weaken
our ability to use empirical data to compare theories.  As Hartle
and Srednicki \cite{hartle+srednicki_07} put it,
\begin{quote}
Cosmological models that predict that at least one instance of
our data exists (with probability one) somewhere in spacetime are
indistinguishable no matter how many other exact copies of these
data exist.
\end{quote}
To illustrate their methods, Hartle and Srednicki introduce a toy
model in which the universe consists of $N$ cycles in time, each
of which is either ``red'' or ``blue'', and each of which
contains an ``observing system'' with probability $p_E$.  They
consider two alternative theories: an ``all red'' ($AR$) theory
in which all cycles are red, and a ``some red'' ($SR$) theory in
which $N_R < N$ cycles are red, and the rest are blue.  Among
other limits, they consider the limit in which $N$ and $N_R$
approach infinity, writing
\begin{quote}
If the number of red cycles is infinite in both models, then the
two theories are not distinguished by our data even if the number
of [{\it sic}] fraction of red cycles is small in SR\null. That
is because our data is that there is at least one red cycle. As
the number of red cycles approaches infinity, that becomes a
certainty in both theories. Even though the typical observing
system in the SR theory is observing blue, our data provides no
evidence that we are typical.
\end{quote}
Thus, we conclude that acceptance of the PRC would make science
as we know it impossible, at least under the assumption that the
universe might be extremely large or infinite. 

Since the PRC seems to be a radical and unmotivated departure
from scientific practice, and since---in the case of a very large
universe---it can render us almost completely powerless to
discover observational reasons to prefer one theory over another,
we see absolutely no reason to adopt it.

\subsection{Rejection of theories in which Boltzmann Brains
dominate}\label{SEC:BBrevisited}

Some of the cosmological theories currently under discussion lead
to very large spacetime regions of exact or approximate de Sitter
space, which leads to a phenomenon known as ``Boltzmann brains''. 
The issue arises from the fact that quantum fluctuations in the
ground state of de Sitter space mimic thermal fluctuations at the
Gibbons-Hawking temperature $T_{\rm GH}=H/(2 \pi)$, where $H$ is
the Hubble expansion rate of the de Sitter
phase~\cite{gibbons+hawking_77}.  In principle, any kind of
thermal fluctuation is possible, with a probability proportional
to $e^{-E/T_{\rm GH}}$, where $E$ is the energy of the
fluctuation (and the Boltzmann constant $k_{\rm B} \equiv1$). 
With some nonzero probability, such thermal fluctuations can give
rise to ``Boltzmann brains'' (BBs), which are objects like our
own brains that are capable of thinking.  Unlike what we believe
for our own brains, BBs are formed as entirely random thermal or
quantum fluctuations, so any memories that they possess are
merely reflections of the random configuration of atoms. 
Nonetheless, with some very small but nonzero probability, a BB
might have exactly the same memories and current thoughts as I
have at this moment.  More generally, there is some nonzero
probability that a BB might be in the same subjective state as
any observer-instant $\,\calI$ allowed by the cosmological
theory.  A BB might be surrounded by a whole Boltzmann planet
arising from thermal fluctuations, or even a whole region
resembling our visible universe.  But for any specified
subjective state $\calS$, the overwhelmingly most likely BB would
be an isolated brain surrounded by a random thermal bath
\cite{rees_97}.  Such fluctuations would be expected to quickly
dissolve into the surrounding thermal equilibrium material,
though even rarer fluctuations could persist for longer, or even
much longer.

We use the term ``ordinary observers,'' or OOs, to describe
observers like we believe we are, who evolved over a period of
$\calO(10^{14})$ years from a low-entropy hot big bang past.  In
some cosmological theories, but by no means all, BBs can
outnumber OOs by an outrageously large or possibly infinite
factor.  We will refer to such theories as ``BB-dominated''.  In
theories of infinite multiverses, the question of whether BBs
dominate over OOs can often depend on the choice of measure that
is adopted to regulate the infinities.  In the scale factor
cutoff measure \cite{desimone+al_10, bousso+al_09}, for example,
the ratio of BBs to OOs is finite and depends on parameters, so
it can be either large or small.  The proper time cutoff measure
does not lead to BB domination, but it is ruled out in any case
by the ``youngness'' problem \cite{linde+al_95, guth_00,
tegmark_05, bousso+al_08}.

As a simple example of a cosmological theory that leads to BB
domination, we can consider the currently standard
$\Lambda$CDM model%
\footnote{That is, the cosmological model which
assumes that the dominant contributions to the energy density of
the universe are a cosmological constant ($\Lambda)$ and cold
dark matter (CDM).} that is used to describe our universe. 
Assuming that the vacuum is absolutely stable, the universe in
this model approaches at late times a future-eternal de Sitter
phase, as the matter density thins to zero, leaving only the
cosmological constant to control the evolution.  The
Gibbons-Hawking temperature for this model is about $T_{\rm GH}
\sim 2 \times 10^{-34}$ eV, so the probability of spontaneously
nucleating anything like a human brain is absurdly small. 
However, since the de Sitter phase would under these assumptions
continue forever, the ratio of BBs to OOs would be infinite.  To
keep the discussion on the topic of very large but finite
universes---to avoid the need to introduce a measure---we will
imagine modifying the description so that the universe ends at
some time, but we will consider an ending time that is
arbitrarily long.  In particular, we will assume that for any
possible subjective state $\calS$, the BBs in that state vastly
outnumber the OOs in that state.

According to the PSLI, we should calculate first-person
probabilities as if our current observer-instant were chosen
randomly and uniformly from all observer-instants in the same
subjective state in the asymptotic ensemble generated from the
theory.  Recall that a real observer in subjective state $\calS$
would have absolutely no way of distinguishing herself from any
of the OOs or BBs in the ensemble in subjective state $\calS$. 
(She could in principle determine from the model universes
$U_\alpha$ which of these observers were OOs and which were BBs,
but since she would not know her own status, there would be no
logic to using this distinction to influence her beliefs about
which of these observers in the model universe describes her own
situation.) Thus, we see no reasonable alternative to her making
predictions as if her current observer-instant were chosen
randomly from this group. 

Using this method, the first-person probabilities would imply
that any real observer is vastly more likely to be a BB than an
OO.  However, even if we focus on BBs who are selected to exactly
match the current subjective state of a real observer, those BBs
would still, with overwhelming probability, see the world around
them rapidly dissolve into a thermal equilibrium mix.  Thus, this
model seems to clearly be in conflict with our observations. 
This consequence of the $\Lambda$CDM model with an absolutely
stable vacuum was noticed by Page \cite{page_08_BB} in 2008, who
suggested that it might show that our vacuum must decay on a time
scale of gigayears---so that BBs cannot dominate. 

But the situation is more subtle than it seems, because if the
true theory really does predict BB domination, then with
overwhelming likelihood we would be BBs.  However, since BBs have
not evolved in coordination with their surroundings over the
history of the universe, everything they think they know is
really just the result of the random configurations of atoms in
their mental circuitry.  There is therefore no reason for a BB to
trust either his memory or thought processes.  So, if BBs really
vastly outnumber OOs, how can we rule out the possibility that we
are BBs who merely think that we can describe and to some extent
understand the real world? It might be all delusion.

Our understanding is that there is no logically sound way to rule
out the possibility that a BB-dominated theory might be correct. 
In that case, however, we would almost certainly be BBs, and we
would have no reason to trust any claims that we can ``do
science''.  That is, there would be no reason to expect any
validity to the theories that we construct, or imagine that we
are constructing.  The conclusion is that there is no point in
pursuing any BB-dominated scientific theory, not because we have
any good reason to believe that they are false, but rather
because they are useless.  If they are valid, then science is
simply not possible.

Is there perhaps still some possibility that the correct theory
is BB-dominated, but we are nonetheless in the situation in which
we happen not to be BBs, and therefore can (we hope) do science?
Such a possibility, we will see, is very strongly disfavored by
the kind of probabilistic reasoning that we have been advocating. 
Suppose, for example, that we consider two theories, $\calT_1$
and $\calT_2$. Theory $\calT_1$ predicts that all observers in my
subjective state, $\calS$, are OOs, while theory $\calT_2$
predicts that only a tiny fraction $q$ of such observers are OOs,
with the rest being BBs.  Suppose that I assign first-person
prior probabilities to the two theories as
$p^{(1p)}(\calT_1|\calS)=p_1$ and $p^{(1p)}(\calT_2|\calS)=p_2 =
1 - p_1$. Now I want to consider the probabilities that ``I can
do science'' and its alternative, ``I cannot do science''.  We
take these statements to be synonymous with ``I am an OO'' and
``I am a BB,'' respectively, and abbreviate them as ``OO'' and
``BB'', respectively.  Using the properties of the theories as
stated, and evaluating first-person probabilities assuming the
PSLI, we find
\begin{equation}
\begin{aligned}
   \pfirst(\calT_1 \hbox{ and OO } | \calS ) &=
      \pfirst(OO|\calT_1, \calS)\, p_1 = p1 \ ,\\
   \pfirst(\calT_2 \hbox{ and OO } | \calS ) &=
      \pfirst(OO|\calT_2, \calS)\, p_2 = q \, p_2 \ ,\\
   \pfirst(\calT_1 \hbox{ and BB } | \calS ) &=
      \pfirst(BB|\calT_1, \calS)\, p_1 = 0 \ ,\\
   \pfirst(\calT_2 \hbox{ and BB } | \calS ) &=
      \pfirst(BB|\calT_2, \calS)\, p_2 = \left(1 - q \right) p_2 \ .
\end{aligned}
\label{EQ:twotheories}
\end{equation}
Assuming that $q \ll p_1/p_2$, which we take as the quantitative
criterion for BB domination, then only two of these four options
have non-negligible probabilities: $\calT_1$ and OO, or $\calT_2$
and BB\null.  Only the first of these two options, however,
allows the possibility for us to do science.  There is no option
in which $\calT_2$ is true {\it and} we can do science.  Thus,
there is no option (with non-negligible probability) in which
$\calT_2$ can be a useful scientific theory.

The subject of Boltzmann brains and their implications has been
controversial in the literature.  Carroll \cite{carroll_17} has
expressed views very similar to ours.  In the abstract, Carroll
wrote ``The issue is not that the existence of such observers [a
preponderance of Boltzmann brains] is ruled out by data, but that
the theories that predict them are cognitively unstable: they
cannot simultaneously be true and justifiably believed.'' We
agree with this conclusion completely, but we have minor quibbles
with some statements that Carroll made along the way.\footnote{In
particular, Carroll wrote (Sec.~5.3) that ``While we know (or at
least seem to perceive) that we are not disembodied brains
floating in an otherwise empty void, we don't know that we and
our local environments haven't fluctuated into existence from a
higher-entropy equilibrium state --- in a long-lived,
randomly-fluctuating universe, it's overwhelmingly likely that we
have indeed done so.'' Although we ``seem to perceive'' that we
are not disembodied brains, we of course cannot trust these
perceptions if we are likely to be Boltzmann brains.  We agree
with Albrecht and Sorbo (in Sec.~III-C of
Ref.~\cite{albrecht+sorbo_04}) that ``The most likely fluctuation
consistent with everything you know is simply your brain
(complete with ``memories'' of the Hubble deep fields, microwave
background data, etc.) fluctuating briefly out of chaos and then
immediately equilibrating back into chaos again.'' Carroll also
wrote (also Sec.~5.3) that ``The best we can do is to decline to
entertain the possibility that the universe is described by a
cognitively unstable theory, by setting our prior for such a
possibility to zero (or at least very close to it). That is what
priors are all about: setting credences for models on the basis
of how simple and reasonable they seem to be before we have
collected any relevant data.'' To us, this sentence seems to
contradict the key point that Carroll and we agree on:
BB-dominated theories should be rejected {\it not} because they
are difficult to believe, but because they cannot be used to do
science.}

In Ref.~\cite{page_17}, Page took issue with Carroll's
Ref.~\cite{carroll_17}, while also, like us, agreeing with
Carroll's main point about cognitive instability.  But Page
argued that BB-dominated theories can be ruled out without
``appeal to this argumentation'' by ``conventional Bayesian
reasoning.'' In our view, Page's argument appeared to rely on the
ability of the reader to understand and verify the Bayesian
argument even if he is a BB, which we find dubious.

Srednicki and Hartle (in Sec.~VII of
Ref.~(\cite{srednicki+hartle_10}) take a very different point of
view, arguing that BB-dominated theories are perfectly
acceptable, as one can prevent the prediction that we should be
BBs by adopting a xerographic distribution that is ``nonzero only
for locations sufficiently close to a big bang to make it much
more likely that we are OOs rather than BBs.'' They argue that
``scientists favor frameworks that are predictive,'' and that a
uniform xerographic distribution would be unpredictive, since
then BBs would dominate, and the observations of any particular
BB would be ``overwhelmingly likely to be disordered, and
inconsistent with its apparent history.''

We agree that predictivity is a very desirable attribute for a
scientific theory, but our view of nonuniform xerographic
distributions is exactly the opposite.  It is of course very
difficult to predict how any living creature will interpret the
stimuli that it receives, and it may be more difficult to do this
for BB's.  But, under our assumption (see Footnote
\ref{FN:Observers}) that observers can be described by the laws
of physics, the observations of BB's are in principle
predictable, like the observations of any other kind of creature.

We argued at the end of Sec.~\ref{SEC:MotivatingIndifference}
that if one takes the liberty to append to a theory an arbitrary
xerographic distribution, the predictivity of the theory
essentially disappears.  If one wishes the theory to predict that
the cosmic microwave background temperature should be 5 degrees,
with everything else matching what we see, one can argue that if
the universe is infinite or sufficiently large, this is likely to
happen somewhere.  We can then choose a xerographic distribution
which requires us to be where it happens, and 5 degrees becomes
the prediction of the theory.  A BB-dominated theory is another
example of this.  Such a theory is unacceptable by our standards,
but like almost any conceivable theory it is acceptable to those
who think it makes scientific sense to imagine a xerographic
distribution that makes us highly atypical.

Gott \cite{gott_08} dismissed Hartle and Srednicki's claims that
we might be atypical, pointing out that
\begin{quote}
All the other authors considering this question have implicitly
assumed the Copernican Principle---or the Principle of Mediocrity
as Vilenkin has sometimes called it. As laid out by Gott
\cite{gott_93} this runs as follows: the Copernican Principle
states that it is unlikely for your location in the universe to
be special among intelligent observers. Why? Because out of all
the places for intelligent observers to be there are by
definition only a few special places and many non-special places,
so you are simply likely to be at one of the many non-special
places.
\end{quote}
This is essentially a version of the counting argument that we
described in Sec.~\ref{SEC:MotivatingIndifference}, although Gott
can be credited with having said it in 1993.  Gott went on,
however, to give two arguments why BB-dominated theories are
nonetheless acceptable.  First, he asserted that BBs are not
intelligent observers, because they cannot pass a Turing test. 
While a very small fraction of BBs can perhaps pass for a human
in answering any specified number of questions, the tendancy of
BBs to be very short-lived would make it highly unlikely that the
next question could be successfully answered.  This argument does
not make sense to us, however, since a Turing test cannot be
infinite.  For any finite number of questions, in a very large
universe the number of BBs that pass the test can vastly exceed
the number of OOs who do.  Gott also argues that BBs are not
real, because they are ``observer dependent.'' This is a subtle
question that involves the emergence of the classical world of
our perceptions from an underlying quantum description, and is
outside the scope of this paper.  We stated in Footnote
\ref{FN:semiclassical} that we do not claim to have a full
understanding of this issue, so we will not comment further on
Gott's argument. 

\subsection{Humans and Jovians}
\label{SEC:Jovians}

In this section we discuss a parable, introduced by Hartle and
Srednicki \cite{hartle+srednicki_07}, that illustrates a
situation which, they claim, shows that reasoning based on the
assumption that we are typical can lead to absurd conclusions. 
Here we argue that their example is indeed absurd, but is by no
means evidence against the PSLI that we defined in
Sec.~\ref{SEC:PSLI}\null.  We will show that the parable can be
modified so that the PSLI applies, and then conclusions are
manifestly sensible.

The parable as described by Hartle and Srednicki goes as follows:

\begin{quote}
{\bf HJ parable}: Consider two theories of the development of
planet-based intelligent life, based on the appropriate physics,
chemistry, biology, and ecology. Theory $A$ predicts that there
are likely to be intelligent beings living in the atmosphere of
Jupiter; theory $B$ predicts that there are no such beings.
Because Jupiter is much larger than Earth, theory $A$ predicts
that there are today many more jovians than humans.

Would we reject theory $A$ solely because humans would not then
be typical of intelligent beings in our solar system? Would we
use this theory to predict that there are no jovians, because
that is the only way we could be typical? Such a conclusion seems
absurd.~\citep[p.~2]{hartle+srednicki_07}
\end{quote}

The absurdity, referred to by the authors, trades on their
surprise about reaching a strong conclusion about the population
on Jupiter by simply demanding that we are typical intelligent
beings (i.e., without making any observations).  Absurd
conclusions are easy to reach if one starts with the blind
assumption that almost anything should be typical of most any
reference group.  As another example, if we assumed that we
should be typical of life on Earth, we would conclude that we are
overwhelmingly likely to be single-cell organisms.  It similarly
sounds absurd to us to assume that life that evolves in the
atmosphere of Jupiter should be similar to intelligent life on
Earth. 

To show that there {\it are} situations in which the assumption
of our typicality can be used rationally to strongly disfavor the
existence of a hypothetical population, we will change two
important characteristics of the HJ parable.  First, Bayesian
logic requires that we are discussing an update based on new
information, and not merely the comparison of two theories {\it
ab initio}.  (To compare two theories {\it ab initio} is to
choose priors, and there are no firm rules about how to do that.)
Second, to apply the PSLI, we must confine the assumption of
typicality to observers in the asymptotic ensemble who are in the
same subjective state. (Only this very limited typicality
assumption is needed to justify the methods proposed in this
paper.)

For the situation to involve an update, we must be initially
ignorant of whether we are ``humans'' or ``jovians'', and to
apply the PSLI it must be possible for us to be in the same
subjective state in either case.  For these reasons, we will
consider a parable that involves only observers on Earth, but the
``jovians'' will be distinguished from the ``humans'' by their
genetics---``jovians'' each have a jovian gene, which is not
linked to any observable traits, and the probability of carrying
this gene is independent of a person's subjective state.  But the
jovian gene can affect the outcome of treatments for one rare
disease, jovian flu.  Like the HJ parable, we will consider two
theories about these populations.  Theory $A$ predicts that the
population of Earth consists of 100 million people with no jovian
gene, and 100 billion people with the jovian gene.  Theory $B$
predicts that the total population is 100 million, and nobody has
the jovian gene.  In the fanciful world of our parable, the
population of the world has not been determined. 

At the start of the parable, nobody knows if they are ``human''
or ``jovian'', and nobody knows whether theory $A$ or theory $B$
is correct.  The update occurs when I come down with jovian flu. 
I see my doctor, and learn more about the disease.  When
untreated, the patient will usually have a very painful rash for
two weeks, but then recover.  Fortunately, however, the hospital
with which my doctor is affiliated has just developed the first
jovian flu antibiotic.  It is completely untested, but reliable
computer simulations show that if the antibiotic is given to an
infected jovian, the patient will be cured in three days, with no
severe symptoms.  However, if the antibiotic is given to a
``human'', it will make the disease worse, leading to a painful
rash for a full month. 

For reasons that he doesn't explain, my doctor assigns a very
strong prior to theory $A$, so he plans my treatment accordingly.
There is, unfortunately, no available test to distinguish
``humans'' from ``jovians'', so the doctor relies on the methods
of statistical medicine.  Since there are 1,000 times as many
``jovians'' as ``humans'' in the general population (according to
his preferred theory, theory $A$), and since he cannot find any
factors in my medical history to suggest that I am different from
the averages, he concludes that the odds of my being a ``jovian''
are 1,000:1.\footnote{Clinicians are trained to look for signs
that their patients might not match the characteristics of a
study group, but in the absence of such signs it is standard
practice to take the percentages of a study group as predicted
probabilities for the patient.  For example, in the {\it
Clinicians' Guide to Statistics for Medical Practice and
Research: Part I}$\,$ \cite{clinicians-guide_06}, an example is
given of a study of heart transplant patients who received hearts
from donors over the age of 35.  27\% of the study patients
experienced cardiac death within 5 years, so the textbook advises
that the data can be used ``to predict that the next patient seen
in clinic \dots\ will have a probability of 0.27 of experiencing
cardiac death within 5 years.'' In other words, in the absence of
evidence to the contrary, it is standard medical practice to
assume that any given patient is typical of the study
group.\label{FN:medical}} He therefore gives me the jovian flu
antibiotic.

Unfortunately, the antibiotic intensifies my disease, and I have
a painful rash for a month before I recover.  Since theory $A$
predicted the probability of this outcome as 1/1001, while theory
$B$ predicted the probability as 1, my doctor used Bayes' theorem
to update his degree of belief in the two theories.  If we denote
his prior estimates as $p_A$ and $p_B$, his posterior
probabilities are found from Bayes' theorem to be
\begin{equation}
\begin{aligned}
p_{{\rm post},A} = \frac{\frac{1}{1001} p_A}{\frac{1}{1001} p_A + p_B}\ ,\\
p_{{\rm post},B} = \frac{p_B}{\frac{1}{1001} p_A + p_B}\ .
\end{aligned}
\end{equation}
That is, theory $A$ has been strongly disfavored, because if
theory $A$ were true, I would have to be significantly atypical. 
Because of this typicality argument, we have strong evidence that
the ``jovian'' population is nonexistent.

The reader will probably have noticed that in our version of the
human-jovian parable, the logic is identical to that used in the
urn problem discussed in Sec.~\ref{SEC:PRC}.  If we draw a white
ball from an urn, we discover evidence that the urn is more
likely to be the one for which white balls are typical.

\subsection{Use of ``Old Evidence''}
\label{APP:OldEv}

This paper has been mainly aimed at developing a method of
assessing theories of a very large universe, in which the
effective, low-energy approximations to the laws of physics might
vary from one region to another.  We have focused on the
definition of first-person probabilities---probabilities
predicted by a given theory for what {\it we} should expect to
see, taking into account all selection effects.  Such
first-person probabilities would then be used to update the
probabilities assigned to theories on the basis of new
experiments, following the usual Bayesian method.  When assessing
theories, however, one cannot avoid the question of how
previously acquired data should affect the probability that we
might assign to a new theory.  Philosophers refer to this as the
`Old Evidence Problem'
\cite{howson_91}.  A commonly discussed example is the role of
the precession of the perihelion of Mercury's orbit as evidence
for general relativity, as the precession was known many years
before the theory was formulated.  The precession of the
perihelion of Mercury is widely viewed as an important piece of
evidence for general relativity, but it does not fit into the
mold of a Bayesian update.  We don't claim to have anything new
to say about this issue, but for completeness we give a brief
summary of the situation as we see it.

In practice, old evidence tends to be evidence that is known to
very good precision, so the assessment of new theories in the
light of old evidence is usually carried out without any
sophisticated statistical analysis.  In looking for alternatives
to general relativity, we limit attention to theories that give
the right prediction for the perihelion precession of Mercury,
within reasonable errors.  Nonetheless, for those of us who have
adopted the Bayesian approach to theory assessment, it seems
important to ask how such old evidence can fit into a Bayesian
analysis.

Many authors \cite{howson_91} (including ourselves) agree that
such old evidence $\calE$ can be used to update the probability
assigned to a theory $\calT$ by imagining that $\calE$ is not yet
known, and then carrying out a standard Bayesian update.  There
are, however, some controversial issues associated with what
exactly is meant by imagining that $\calE$ is not yet known,
since $\calE$ might be logically intertwined with other pieces of
information in our knowledge base.\footnote{Ref.~\cite{howson_91}
gives an example in which our knowledge base contains the pair of
propositions $\{\calE, \calE \rightarrow \calB\}$. But the
knowledge base might alternatively be described by including the
pair of propositions $\{\calE,\calB\}$, which leads to the same
consequences. The two descriptions are equivalent, but if $\calE$
is removed, then $\calB$ would still hold in the second case, but
not the first.}

To resolve this issue, we would suggest that the situation in
which $\calE$ is not yet known can be defined historically.  We
can imagine turning back the clock to the time before $\calE$ was
discovered.  Note that by turning back the clock, we are
presumably enlarging the reference class, since it now includes
all observer-instants that share all memories up to some point in
the past, but which might diverge for later times.  To carry out
this turning back of the clock, we would have to ask what would
have been reasonable priors to assign to the relevant theories at
that time, and then we could carry out the Bayesian update.  In
choosing priors, we would be using modern theories, with a modern
understanding of their consequences, but we would need to rely
exclusively on experimental information available at the earlier
time.  Such a process would produce probability assignments that
would have been relevant just after $\calE$ was discovered, so to
obtain currently relevant probabilities we might have to continue
the process to include all other relevant experiments between
then and now.  A crucial difficulty with this procedure is the
assignment of prior probabilities at the time just before $\calE$
was discovered, because it is hard to know how much our choices
might be biased by knowledge of data obtained later.  For that
reason, the implications of old evidence are generally more
controversial---more subject to the differing opinions of various
researchers---than the implications of new experiments.

\section{Conclusion}
\label{SEC:Conclusion}

Probabilistic reasoning in very large universes is a thorny
issue, and analyses of these novel physical settings have led to
confusion and controversy in the literature.  We are interested
in theories that describe universes in which the effective,
low-energy approximations to the laws of physics can vary from
one region to another, so the measurements of any observer will
strongly depend on where she lives.  Furthermore, the local
low-energy laws of physics in any region can have a strong
influence on whether observers exist in the region.  If observers
do exist, their number and their characteristics will also be
strongly influenced by the local low-energy laws.  In this
complicated situation, there is as yet no generally accepted
procedure to calculate probabilities for what a given observer
might experience, or for how she should assess the credibility of
competing theories.

In this paper we have argued that these questions are not as
novel as some authors have claimed, and that the best method of
analysis is closely analogous to standard methods of physics.  We
distinguish between third-person probabilities, that would be
measured by a hypothetical external observer who has access to
the entire universe, and first-person probabilities, that would
be measured by any particular real observer within the universe,
taking into account all selection effects associated with the
observer.  We have made a specific proposal for how any
particular observer should calculate the first-person
probabilities for what she will experience.We proposed that such
first-person probabilities should be used to update probabilities
assigned to competing theories, using the completely standard
Bayesian procedure.  We focused on the principles that define
this method, ignoring any practical issues associated with
implementing it.  We believe that it is important to settle these
issues of principle first, and later one can seek practical
approximations.

We described a generic theory $\calT$ as a system that provides a
set of possible universes $\{U_\alpha\}$, and a probability
$\Pcos(U_\alpha)$ that any one of them describes the real
universe $\Ureal$.  Each $U_\alpha$ is a complete description of
the entire spacetime, and everything in it.  We then imagined
constructing a theory-generated ensemble of possible universes,
with each possible universe contributing to the ensemble in
proportion to $\Pcos(U_\alpha)$.

To make predictions for any particular observer $\calO$, one must
take into account all selection effects.  We argued that in
principle each observer is subject to her own selection effects,
and that these selection effects can in general depend on time. 
We concluded that predictions should be made separately for each
observer-instant $\,\calI$, referring to a particular observer
$\calO$ at a particular time $t$.  Predictions can be made by
associating a given observer-instant $\,\calI$ in the real
universe with a reference class of observer-instants $\,\calI_i$
in the theory-generated ensemble.  We believe that the best
method of prediction should take into account all available
information, so we defined the {\it subjective state} $\calS$ of
an observer-instant as the set of all properties that the
observer believes, at time $t$, about herself and her
surroundings.  The reference class then consists of all observer
instants $\,\calI_i$ in the same subjective state as $\,\calI$. 

There will in general be many different $\,\calI_i$'s in
different model universes $U_\alpha$ in the ensemble, so we need
a procedure to extract predictions from the set.  By these
definitions the real observer $\calO$ cannot, at time $t$,
distinguish any of the observer-instants $\,\calI_i$ from her own
observer-instant, so she has no evidence that any one of them is
more likely to describe her future than any other.  These are
precisely the conditions under which the principle of
indifference is usually taken to apply.  In this context we
propose a {\it Principle of Self-Locating Indifference} (PSLI),
which holds that an observer should predict first-person
probabilities for her future as if her current observer-instant
were chosen randomly and uniformly from the observer-instants in
the ensemble that are in the same subjective state.  We pointed
out that this is analogous what physicists assume in standard
laboratory settings, where it is always assumed that the results
obtained in the experiment are typical, within expected
statistical fluctuations, of the results that any experiment of
the same type would yield.

We believe that the PSLI is a very reasonable principle, but we
also argue that using this method optimizes the accuracy of
predictions.  In a probabilistic system there is never any way to
guarantee that a specific prediction will turn out to be true,
but we argued (in Appendix~\ref{APP:PSLIknown}) that the use of
the PSLI by real observers maximizes the fraction of them who
will make correct predictions if the theory is correct. Whether
one is discussing the outcomes for the flipping of idealized fair
coins or for experiments in a very large universe, one can
predict what most observers will find, but one cannot guarantee
what any particular observer will find.  In either case, we
believe that it is sensible to assume that what most observers
will find is most likely what I will find.  Some authors have
argued that there is a ``random process'' that motivates this
assumption for the case of coin flipping, but not for the
self-locating uncertainties involved in predicting experimental
outcomes in a very large universe.  In Sec.~\ref{SEC:Fallacy}, we
argued that there is no relevant difference between these two
situations.

A possibly controversial consequence of the PSLI is the
``weighting by the number of observers''---that is, in
determining first-person probabilities, the weight associated
with any possible model universe $U_\alpha$ is proportional to
both its third-person probability $\Pcos(U_\alpha)$ and to the
number of observer-instants it contains that are in the same
subjective state as the observer-instant for which predictions
are sought.  In contrast, Hartle and collaborators (in
Refs.~\cite{hartle+srednicki_07, srednicki+hartle_10,
hartle+hertog_15, hartle+hertog_16, hartle+al_10}, for example)
advocate weighting each possible model universe only by its
third-person probability, restricted to universes that have at
least one instance of our data $D$.\footnote{The definition of
``data'' used by Hartle and collaborators seems very similar to
our definition of ``subjective state''.  In
Ref.~\cite{srednicki+hartle_10}, the authors describe ``the data
$D_0$ that the HSI [human scientific IGUS, where IGUS =
information gathering and utilizing system] has: every scrap of
information that the HSI possesses about the physical universe,
including every record of every experiment, every astronomical
observation of distant galaxies, every available description of
every leaf, etc., and necessarily every piece of information
about the HSI itself, its members, and its history.''} In
Appendix~\ref{APP:simp_assump}, we show how the weighting by
observers can be derived from a set of simpler, reasonable
assumptions. 

One issue for which probabilistic reasoning in very large
universes is relevant is the question of ``Boltzmann brains''. 
In Sec.~\ref{SEC:BBrevisited}, we argued that theories in which
Boltzmann brains dominate over ordinary observers should be
rejected.  They should be rejected not because we see a rational
reason why they cannot be valid, but instead because, if they are
valid, then we would most likely be Boltzmann brains and would
not be capable of doing science.  Thus we see no possibility that
such theories can be useful for scientific investigations.

One possibly significant issue that we did not consider in this
paper, but leave as a topic for future investigation, is the time
evolution of subjective states and its possible relevance to the
calculation of first-person probabilities.  One possibility is
that subjective states are well-modeled as being discrete, so an
observer evolves for a short period of time in one subjective
state, and then transitions instantaneously to a subsequent
state.  In that case, it might be appropriate to calculate
first-person probabilities by weighting these intervals in
proportion to their duration, rather than assuming that all
observer-instants count equally.  Alternatively, it might be
better to treat the time evolution of subjective states as a
continuous evolution, in which case one might consider whether
the weight given to an observer-instant might depend on how
quickly the observer passes through the specified subjective
state.

The methods developed in this paper are relevant to several
interesting questions which we have not discussed here. One is
the question of whether evidence that the universe is fine-tuned
to allow for the possibility of life should be interpreted as
evidence in favor of a multiverse.  Another is the so-called
``Doomsday Argument''---the assertion that human civilization is
not likely to survive much longer than its current age, since a
much longer period of survival would place us atypically early in
human history.  We leave these questions as topics for future
research.

Note that this paper is directly relevant to very large universes
where the total spacetime volume is finite.  However, many
cosmological theories under consideration today describe infinite
``multiverses,'' where the definition of probabilities requires a
solution to what is known as the measure problem.  Of course
there remains much work to be done.  We are very far from being
able to produce cosmological theories for which we can calculate
consequences at the level assumed in this paper.  It is not clear
how the measure problem will be resolved, but perhaps theoretical
reasoning can lead to a unique solution, or perhaps it will be
necessary to consider a range of solutions to be tested
observationally.  In any case, we expect that the methods
developed in this paper, based on the {\it Principle of
Self-Locating Indifference}, will continue to be applicable as
these advances unfold. 

\begin{acknowledgments}
We would like to thank Jeremy Butterfield, Don Page, and Scott
Sheffield for helpful conversations. F.~A.~acknowledges support
from (i) the Black Hole Initiative at Harvard University, where
this work began, which is funded through a grant from the John
Templeton Foundation and the Gordon and Betty Moore Foundation;
and (ii) the Faculty Research Support Program (FY20FRSPE) at the
University of Notre Dame. A.H.G.'s work was supported in part by
the U.S. Department of Energy, Office of Science, Office of High
Energy Physics under grant Contract Number DE-SC0012567.
\end{acknowledgments}

\appendix

\addcontentsline{toc}{section}{{\hskip -3.6111pt}Appendices:}

\section{Motivating the \textit{Principle of Self-Locating Indifference}
assuming that the model $U_*$ for the real universe is known}
\label{APP:PSLIknown}

As part of our justification of the {\it Principle of
Self-Locating Indifference}, here we adopt the simplifying
assumption that we know which model universe $U_* \in
\{U_\alpha\}$ is actually realized.  Since the goal is to define
a procedure for extracting first-person predictions from theories
in very large universes, we will use the success of such
predictions as our criterion for comparing different possible
procedures. Thus, we will motivate the PSLI by considering a toy
model of an experiment, and we will show that the expected number
of observers who will make correct inferences from their
experiments is maximized if every observer interprets his data
using the PSLI. This is an application of the same ``counting
argument'' that we discussed in
Sec.~\ref{SEC:MotivatingIndifference}: if we can devise a
procedure for which observers who make incorrect inferences are
very rare, then we can assume that we are very unlikely to fall
into this class.

For our toy model experiment, we will imagine that some
observable quantity $\theta$ is measured by many different
observer-instants $\,\calI$ in the same subjective state, all in
the universe described by $U_*$. The quantity $\theta$ might be,
for example, the value of the Higgs mass, or the fine-structure
constant, or the cosmological constant.  While any observable in
reality could be correlated with any number of other observables,
we will assume for our toy model that $\theta$ has no
correlations with any other observable.  Furthermore, we will
assume that $\theta$ has the same value throughout the universe. 

Consider two possible assumptions about typicality that each of
the observers could adopt: `typicality' and `atypicality'. Define
`typicality' as the assumption of the PSLI---the assumption that
each observer should calculate first-person probabilities as if
their current observer instant were chosen randomly and uniformly
from among all the observer-instants $\calI_i$. Observers who
adopt `typicality' will be referred to as `typicality-accepting'.
We also consider an `atypicality' assumption---the assumption
that each observer should calculate first-person probabilities as
if their current observer instant is expected to be atypical in
some specific way. Observers who adopt `atypicality' will be
referred to as `typicality-denying'. For our toy model,
`typicality-denying' observers will assume that their
first-person expectation value for the measurement of $\theta$ is
higher than the theoretical value by $10\%$. (Note that the
assumption that the first-person probability distribution for
$\theta$ is affected by the atypicality assumption is generic. If
no observables are affected, then the atypicality assumption is
completely irrelevant. As long as at least one observable is
affected, we can choose $\theta$ to be that observable. We assume
here for simplicity that the expectation value is affected, but
similar conclusions could be justified if the standard deviation
or any other property of the probability distribution is
affected.)

The observers use their assumptions about their own typicality in
interpreting the results of their measurements. We will show that
the assumption of typicality leads to the largest expectation
value for the fraction of observers reaching accurate inferences.
In the absence of any way to know where we will lie in the
distribution of outcomes, the best way to ensure that our
inferences will be correct is to maximize the expectation value
of the number of observers who will make accurate inferences. 

The observers have (the same) theories of the measurement of this
observable at their disposal.  Each theory yields a probability
density function for the measurement of $\theta$, which includes
any experimental errors that arise in the measurement process, as
well as any quantum uncertainties intrinsic to the measurement.
We denote a given theory by $\mathcal{T}(\theta_t)$, where we
have labeled the theory via the mean, $\theta_t$, of the
probability density function that the theory yields for $\theta$.
The probability density function for measuring $\theta$,
according to the theory $\mathcal{T}(\theta_t)$, will then be
denoted by $p(\theta|\theta_t)$.

We will proceed under the simplifying assumption that this
distribution is a normal distribution,
\begin{equation}
  p(\theta|\theta_t) = 
  \mathcal{N}_{\theta}(\theta_t, \sigma), 
\end{equation}
where $\mathcal{N}_{\theta}(\theta_t, \sigma)$ represents a
normal distribution in $\theta$ with mean $\theta_t$ and standard
deviation $\sigma$. Denote the true value of $\theta$ in the
model universe $U_*$ by $\theta_{\rm true}$, which we assume to
be nonzero.  For our toy model experiment, we will take $\sigma =
\theta_{\rm true}/100$.  We will denote the measured value of
$\theta$ by $\theta_D$ (where ``$D$'' is for ``data''). 

For our example, suppose that all the observers adopt a prior
probability distribution that is flat and ranges from
$-10\theta_{\rm true}$ to $10\theta_{\rm true}$; that is:
\begin{equation}\label{EQ:Unif} 
p_{\rm prior}(\theta_t) = U_{\theta_t}\left(-10\theta_{\rm true},10\theta_{\rm true} \right),
\end{equation}
where $U_{\theta}(a, b)$ denotes the uniform distribution for the
random variable $\theta$, between $\theta = a$ and $\theta = b$.
[The observers, of course, do not know the value of $\theta_{\rm
true}$, but we will assume for our example that their choice of
prior corresponds to Eq.~(\ref{EQ:Unif}).]

According to the assumptions stated above about
`typicality-accepting' ($a={\rm typ}$) and `typicality-denying'
($a={\rm atyp}$) observers, the first-person probabilities that
these observers will adopt can be written as
\begin{equation}\label{EQN:a1norm}
p^{(1p)}_a({\theta}|\theta_t)\equiv \mathcal{N}_{{\theta}}
  \left(\kappa_a \theta_{\rm t}, {\sigma}\right), 
\end{equation}
where
\begin{align}
	\kappa_a = 
	\begin{cases} 
     1 & \hbox{if } a = {\rm typ} \\
     1.1 & \hbox{if } a = {\rm atyp}.
	\end{cases}
\end{align}

Recall that we are working in a Bayesian setting, where the
primary distribution of interest, as the observer completes
further trials, is the first-person posterior probability
$p^{(1p)}_{{\rm post}; a}(\theta_t|{\theta_D})$. This is given,
in general, by Bayes' theorem:
\begin{equation}\label{EQN:a1post}
p^{(1p)}_{{\rm post}; a}(\theta_t|\theta_D) = \frac{p^{(1p)}_a(\theta_D|\theta_t)\,p_{\rm prior}(\theta_t)}{p(\theta_D)},
\end{equation}
where
\begin{equation}\label{EQN:a1LTP}
p(\theta_D) = \int\,d\theta_t\, p^{(1p)}_a(\theta_D|\theta_t)\, p_{\rm prior}(\theta_t). 
\end{equation}
Our two classes of observers can then compute two different
first-person posterior probability distributions for $\theta_t$
by substituting Eqs.~(\ref{EQ:Unif}) and (\ref{EQN:a1norm}) into
Eq.~(\ref{EQN:a1post}).

The posterior probability density for either kind of observer
peaks at $\theta_t = {\theta_D}/\kappa_a$. One may ask, what is
the probability that, according to the true theory, the peak is
within a range around the true value, say in the range
$[\theta_{\rm{true}}(1-\epsilon),
\theta_{\rm{true}}(1+\epsilon) ]$ where $\epsilon$ is a small
number. This probability is given by
\begin{widetext}
\begin{align}
P\Bigg(\left|
     (\theta_D/\kappa_a)
     - \theta_{\rm{true}}\right|<
     \epsilon\,\theta_{\rm{true}}
     \, \bigg|
     \,\theta_{\rm{true}}\Bigg) & =
     \int_{\kappa_a \theta_{\rm{true}}
     (1-\epsilon)}^{\kappa_a\theta_{\rm{true}}
     (1+\epsilon)} p(\theta_D |\theta_{\rm{true}}) d \theta_D
     \nonumber \\
& =
\frac12 \text{erf}\left(\frac{ \theta_\text{true} }{\sqrt{2}
     \sigma } 
     [\kappa(1 + \epsilon) - 1]
     \right)
     -
     \frac12 \text{erf}\left(\frac{ \theta_\text{true}}{\sqrt{2}
     \sigma } 
     [\kappa(1-\epsilon) - 1]
     \right)\, .
\end{align}
\end{widetext}

For `typicality-accepting' observers, where $\kappa_{\rm typ} =
1$, this probability is significantly higher than for
`typicality-denying' observers. For example, for $\epsilon=2\%$,
this probability is 0.95 for `typicality-accepting' observers and
$3.1\times 10^{-15}$ for `typicality-denying' observers. Thus the
expected number of observers who will find the peak in the
specified range is vastly larger if the observers are
`typicality-accepting'.

The toy model example described here is very simplified, and uses
numbers chosen to be fairly extreme, but the qualitative result
is universal.  Ideal experiments are designed to be ``unbiased,''
which means that the expectation value of the result should be
equal to the true value of the quantity.  The expectation value
is what, in our language, is the expected average value that
would be obtained by typical observers.  If one were to apply a
correction to the result, motivated by the assumption that we are
atypical, then the protocol would become biased.  The toy model
example in this appendix is really just a step-by-step
illustration of why unbiased measurements are considered to be
obviously preferable to biased measurements (provided that all
other aspects are equal).

\section{Motivating the \textit{Principle of Self-Locating Indifference}:
Dropping the assumption that $U_*$ is known}
\label{APP:simp_assump}

In this appendix we continue our justification of the {\it
Principle of Self-Locating Indifference}, now dropping the
assumption that we know which $U_\alpha$ is the model universe
$U_*$ that is realized---that is, which model universe $U_\alpha$
describes the real universe $\Ureal$.  This requires us to show
that, for any given observer-instant $\calI$ in the real world,
the probability that any proposition $\calP$ is true can be
calculated as if $\calI$ were chosen randomly and uniformly from
the set of all observer-instants in the asymptotic ensemble that
are in the same subjective state $\calS$.  In Sec.~\ref{SEC:PSLI}
we showed that this statement is equivalent to evaluating
$\pfirst(\calP | \calS)$, first-person probability for a
proposition $\calP$, given $\calS$, using Eq.~(\ref{EQN:P1pP}). 
In this Appendix we will derive Eq.~(\ref{EQN:P1pP}) starting
from assumptions that we consider very reasonable, and somewhat
more primitive than the assumption of the PSLI itself.

One of the most striking features of the PSLI, which will be
justified in this appendix, is the weighting by the number of
observers.  That is, the PSLI implies that
$\pfirst(U_\alpha|\calS)$, the first-person probability that a
given observer-instant $\calI$ will find itself in a universe
described by $U_\alpha$, is weighted by the number of
observer-instants found in $U_\alpha$ that are in the same
subjective state $\calS$ that $\,\calI$ is.

Throughout this appendix we will be concerned with defining
first-person probabilities for a particular observer-instant
$\,\calI$ {\it in the real universe}.  The subjective state of
$\,\calI$ will be denoted by $\calS$. 

For each model universe $U_\alpha$ allowed by the theory, there
might be any number of observer-instants $\,\calI_i$ that are in
the subjective state $\calS$.  Although they are all in the same
subjective state, they all have different spacetime locations, so
each can be labeled with a distinct index $i$, as we assumed when
we first wrote ``$\calI_i$''.  We will refer to $i$ as the {\it
observer-instant number}, which ranges from $i=1$ up to the
number of such observer-instants in $U_\alpha$, which we denote
by $\ntot(U_\alpha, \calS)$.  It does not matter how the
observer-instant numbers are assigned, so we can imagine that
they are assigned randomly by whoever is analyzing the theory to
extract its predictions. 

Our discussion will be based on the asymptotic ensemble derived
from the theory under investigation, as described in detail in
Sec.~\ref{SEC:PSLI}\null.  Each entry $\Uens_\lambda$ is one of
the possible universes $U_\alpha$, so each observer-instant in
$\Uens_\lambda$ can be assigned the observer-instant number that
it inherits from $U_\alpha$.  Thus, each observer-instant in the
ensemble can be labeled by the triplet $(\lambda, \calS, i)$,
where $\Uens_\lambda$ is the model universe in which it occurs,
$\calS$ is its subjective state, and $i$ is its observer-instant
number. 

To proceed, we will argue that in the context of extracting
first-person probabilities from a given theory, we can imagine
that each observer-instant {\it in the real universe} can also be
assigned a unique observer-instant number.  To see this, first
note that the first-person probability of any proposition $\calP$
implied by a given theory is simply the answer to the question
``if the theory is correct, what is the first-person probability
of $\calP$?''.  That is, when extracting a predicted probability
from a theory, one is always conditionalizing on the assumption
that the theory is correct.  Under that assumption, the real
universe $\Ureal$ must be perfectly described by some $U_* \in
\{U_\alpha\}$, where $\{U_\alpha\}$ is the set of possible
universes allowed by the theory.  While the subjective state
$\calS$ of the observer-instant $\,\calI$ in the real universe
may occur many times in $U_*$, the spacetime location of
$\,\calI$ occurs only once, so $\,\calI$ can in principle be
identified with a particular $\,\calI_i$ in $U_*$, and can
therefore be assigned the observer-instant number $i$.  The
observer-instant number for $\,\calI$ can be determined only by
an external observer who has access to the entire real universe,
but we see no barrier to considering it in probabilistic
calculations. 

Suppose that we know that $i$ is the observer-instant number for
$\,\calI$, but we do not know which possible universe is the one
that is realized.  In that case the mere existence of
observer-instant number $i$ (for the subjective state $\calS$)
implies that
\begin{equation}
   n_{\rm tot}(U_\alpha, \calS) \ge i \ .
\label{eq:nfits}
\end{equation} 
The criterion of Eq.~(\ref{eq:nfits}) divides the entries of the
ensemble into two classes: (1) {\it allowed universes}, for which
Eq.~(\ref{eq:nfits}) holds, and which therefore contain exactly
one observer-instant compatible with $\,\calI$, and (2) {\it
forbidden universes}, for which the equation fails, and which
therefore cannot be the realized universe. 

Finally, to determine the first-person probability
$\pfirst(U_\alpha, i|\calS)$, we will make two assumptions:.

\begin{itemize}

	\item[$\calA_1$:]
     {Suppose that we know which model universe $U_* \in
     \{U_\alpha\}$ describes the real universe $\Ureal$, as we
     assumed in the previous appendix.  Then
     first-person
     predictions for $\,\calI$ should be made as if $\,\calI$
     were chosen randomly and uniformly from the set of observer
     instants $\,\calI_i$ in $U_*$ in the same subjective state,
     which implies that $i$ should be chosen randomly and
     uniformly from the set $\{1,2,\ldots,\ntot(U_*,\calS)\}$. 
     (This statement is just the PSLI restricted to the case in
     which $U_*$ is known, as discussed in the previous
     appendix.)}

	\item[$\calA_2$:]
     {Suppose that we know the observer-instant number $i$ of
     observer-instant $\,\calI$, but we don't know which model
     universe $U_* \in \{U_\alpha\}$ describes the real universe.
     Then first-person predictions for $U_*$ should be made as if
     $U_*$ were chosen randomly and uniformly from the {\it
     allowed} universes in the asymptotic ensemble.}
\end{itemize}

To proceed, we need to consider the realistic case in which we do
not know either $U_*$ or $i$.  Thus, we focus on
$\pfirst(U_\alpha, i | \calS)$, the first-person conditional
probability that $U_\alpha$ is the universe that is realized and
that $i$ is the observer-instant number of $\,\calI$, given the
subjective state $\calS$.  We will find that $\calA_1$ determines
the dependence of this quantity on $i$, and that $\calA_2$
determines its dependence on $U_\alpha$, so it is determined
completely. 

Before giving the general proof, we will give a simple example to
illustrate the logic.  Suppose that a theory predicts only two
possible universes, $U_1$ and $U_2$, with equal probabilities
$\Pcos(U_\alpha)$, where $U_1$ contains a million
observer-instants in my subjective state $\calS$, while $U_2$
contains only one.  Now suppose that I am told by a hypothetical
external observer that my observer-instant number is 1.  In that
case, there is exactly one observer-instant in $U_1$ compatible
with my subjective state and observer-instant number, and
similarly one compatible observer-instant in $U_2$.  Both
universes are therefore allowed, so by assumption $\calA_2$, I am
equally likely to live in either universe:
$\pfirst(U_2,i\hbox{=}1|\calS) =
\pfirst(U_1,i\hbox{=}1|\calS)$.  But according to assumption
$\calA_1$, if I live in universe $U_1$, I should be equally
likely to have any observer-instant number from 1 to 1,000,000. 
So $\pfirst(U_1,i|\calS) = \pfirst(U_1,1|\calS)$, for $i \in
[1,10^6]$.  It then follows that if I don't know my
observer-instant number, the first-person probability that I live
in $U_1$ is given by $\pfirst(U_1|\calS) = \sum_i
\pfirst(U_1,i|\calS) = 10^6\,
\pfirst(U_1,i\hbox{=}1|\calS)
= 10^6\,
\pfirst(U_2,i\hbox{=}1|\calS)
= 10^6\,
\pfirst(U_2|\calS)$,
demonstrating the weighting by the number of observers.

To construct the general proof, we begin by expressing assumption
$\calA_1$ by the equation
\begin{equation}
\pfirst(i|U_\alpha,\calS) = \frac{\Theta(U_\alpha,\calS,i)}
{n_{\rm tot}(U_\alpha,\calS)} \ ,
\label{eq:A1-in-eq}
\end{equation}
where
\begin{equation}
\Theta(U_\alpha, \calS, i) = \begin{cases} 
1 & \hbox{if } n_{\rm tot}(U_\alpha, \calS) \ge i \\
0 & \hbox{otherwise,}
\end{cases}
\end{equation}
and where we are assuming that $n_{\rm tot}(U_\alpha,\calS) > 0$. 
If $n_{\rm tot}(U_\alpha,\calS) = 0$ and observer-instant
$\,\calI$ exists, then the combination $(U_\alpha, \calS)$ is not
possible. Using the definition of conditional probability,
$\pfirst(i|U_\alpha,\calS) = \pfirst(U_\alpha,i |
\calS)/\pfirst(U_\alpha|\calS)$, Eq.~(\ref{eq:A1-in-eq}) can be
rewritten as
\begin{equation}
\pfirst(U_\alpha, i|\calS) = \frac{\pfirst(U_\alpha|\calS)
     \Theta(U_\alpha,\calS,i)} {\ntot(U_\alpha,\calS)} \ . 
\label{eq:ndependence}
\end{equation}

To extract the consequences of $\calA_2$, recall that choosing a
universe randomly and uniformly from the asymptotic ensemble is
equivalent to choosing it according to the theory-generated
probability $\Pcos(U_\alpha)$.  Thus, $\calA_2$ can be expressed
as the equation
\begin{subequations}
\begin{align}
\pfirst(U_\alpha|\calS, i) &= p\bigl(U_\alpha
     | U_\alpha \hbox{ is allowed for }(\calS,i) \bigr)
\\
&= p\Bigl(U_\alpha \Big| \, \Theta(U_\alpha,\calS,i) = 1\Bigr) \\
&= \frac{1}{Z_1(\calS,i)} p\Bigl(U_\alpha \hbox{ and }
     \Theta(U_\alpha,\calS,i) = 1\Bigr) 
     \label{EQ:A2-eqb}\\
&= \frac{1}{Z_1(\calS,i)} \Pcos(U_\alpha)
     \Theta(U_\alpha,\calS,i) \ ,
     \label{EQ:A2-eqc}
\end{align}
\end{subequations}
where
\begin{equation}
Z_1(\calS,i) = \sum_\beta \Pcos(U_\beta) \Theta(U_\beta, \calS,
     i) \ .
\end{equation}
To understand why the expression (\ref{EQ:A2-eqc}) is equal to
the expression (\ref{EQ:A2-eqb}), note that $\Theta(U_\alpha,
\calS, i)$ is either equal to 1 or 0, and that the equality holds
in both cases. 

By the definition of conditional probability,
$\pfirst(U_\alpha,i|\calS) = \pfirst(U_\alpha|\calS,i)
Z_2(\calS,i)$, where $Z_2(\calS,i) \equiv \pfirst(i|\calS)$, so
with Eq.~(\ref{EQ:A2-eqc}) we have
\begin{equation}
\pfirst(U_\alpha,i|\calS) = \Pcos(U_\alpha) \Theta(U_\alpha,
     \calS, i) \frac{Z_2(\calS,i)}{Z_1(\calS,i)} \ .
\label{EQ:idependence}
\end{equation}

To combine the information in Eqs.~(\ref{eq:ndependence}) and
(\ref{EQ:idependence}), note that when
$\Theta(U_\alpha,\calS,i)=0$, both equations imply that
$\pfirst(U_\alpha|\calS, i) = 0$.  When
$\Theta(U_\alpha,\calS,i)=1$, their equality implies that
\begin{equation}
\frac{\pfirst(U_\alpha|\calS)} {\ntot(U_\alpha,\calS)} =
     \Pcos(U_\alpha) \frac{Z_2(\calS,i)}{Z_1(\calS,i)} \ ,
\label{EQ:pfirst-alpha-1}
\end{equation}
Since the left-hand side is independent of $i$, the right-hand
side must also be, so we can write
\begin{equation}
\frac{Z_2(\calS,i)}{Z_1(\calS,i)} = \frac{1}{Z_3(\calS)}
\label{EQ:Z3}
\end{equation}
for some function $Z_3(\calS)$. 

One can also insert Eqs.~(\ref{EQ:pfirst-alpha-1}) and
(\ref{EQ:Z3}) into Eq.~(\ref{eq:ndependence}), showing that
\begin{equation}
\pfirst(U_\alpha, i|\calS) = \frac{1}{Z_3(\calS)} \Pcos(U_\alpha)
     \Theta(U_\alpha, \calS, i) \ .
\label{EQ:alpha-i}
\end{equation}
The above equation was derived under the assumption that
$\Theta(U_\alpha, \calS, i) = 1$, but it is clearly also true in
the alternative case, where $\Theta(U_\alpha, \calS, i) = 0$ and
both sides are zero. 

$Z_3(\calS)$ can be determined by normalization:
\begin{equation}
\begin{aligned}
Z_3(\calS) &= \sum_{\alpha,i} \Pcos(U_\alpha) \, \Theta(U_\alpha,
     \calS, i) \\ 
   &= \sum_\alpha \Pcos(U_\alpha) \ntot(U_\alpha,\calS) \ .
\label{EQ:Z3-2}
\end{aligned}
\end{equation}

To rederive Eq.~(\ref{EQN:P1pP}), we can use
Eq.~(\ref{EQ:alpha-i}) to calculate the first-person probability
for any proposition $\calP$, given the subjective state $\calS$
of the observer-instant $\calI$:
\begin{equation}
  \pfirst(\calP | \calS) = \sum_{\alpha,i} \pfirst(\calP |
     U_\alpha, \calS, i) \, \pfirst(U_\alpha,i | \calS) \ ,
     \label{EQ:alpha-i-P}
\end{equation}
where $\pfirst(\calP | U_\alpha, \calS, i)$ is the probability
that the proposition $\calP$ is true for a specified
observer-instant, given the model universe $U_\alpha$, the
subjective state $\calS$, and observer-instant number $i$ of the
observer-instant. Since the observer-instant is fully described
by these parameters, $\calP$ is either true or false for this
observer-instant, so
\begin{equation}
  \pfirst(\calP | U_\alpha, \calS, i) = \begin{cases}
     \begin{aligned} & 1 \hbox{ if } \calP \hbox{ is true for
           observer-instant } \\
        & \qquad (U_\alpha, \calS, i) 
     \end{aligned} \\
   0 \hbox{ otherwise.}
   \end{cases}
\end{equation}
Substituting $\pfirst(U_\alpha,i | \calS)$ from
Eq.~(\ref{EQ:alpha-i}) into Eq.~(\ref{EQ:alpha-i-P}),
\begin{equation}
\begin{aligned}
  \pfirst(\calP | \calS) &= \frac{1}{Z_3(\calS)} \sum_{\alpha,i}
     \Pcos(U_\alpha) \, \Theta(U_\alpha, \calS, i) \\
     & \qquad \times \pfirst(\calP | U_\alpha, \calS, i) \ .
  \label{EQ:alpha-i-P-2}
\end{aligned}
\end{equation}
Note that whenever $\pfirst(\calP | U_\alpha, \calS, i)=1$, the
proposition $\calP$ must be true for the observer-instant
indicated by $(U_\alpha, \calS, i)$, and therefore the
observer-instant must exist, and therefore $\Theta(U_\alpha,
\calS, i)=1$.  Thus, the factor $\Theta(U_\alpha, \calS, i)$ in
the above equation is superfluous and can be dropped:
\begin{equation}
  \pfirst(\calP | \calS) = \frac{1}{Z_3(\calS)} \sum_{\alpha,i}
     \Pcos(U_\alpha) 
     \,\pfirst(\calP | U_\alpha, \calS, i)
     \ . 
  \label{EQ:alpha-i-P-3}
\end{equation}
Now note that
\begin{equation}
  \sum_i 
     \pfirst(\calP | U_\alpha, \calS, i)
     = \nprop(U_\alpha, \calS, \calP) \ ,
  \label{EQ:nprop}
\end{equation}
where we recall that $\nprop(U_\alpha, \calS, \calP)$ was defined
in Sec.~\ref{SEC:PSLI} as the number of observer-instants in
subjective state $\calS$ in the model universe $U_\alpha$ for
which the proposition $\calP$ is true.  Using
Eqs.~(\ref{EQ:alpha-i-P-3}), (\ref{EQ:nprop}), and
(\ref{EQ:Z3-2}), we have finally
\begin{equation}
  \pfirst(\calP | \calS) = \frac{\displaystyle \sum_{\alpha}
     \Pcos(U_\alpha) \nprop(U_\alpha, \calS,
     \calP)}{\displaystyle \sum_\beta \Pcos(U_\beta) \,
     \ntot(U_\beta,\calS)} \ ,
  \label{EQ:P1pP2}
\end{equation}
which is identical to Eq.~(\ref{EQN:P1pP}).  Thus,
we have rederived the PSLI.%
\footnote{The reader may have noticed that the
situation discussed here has analogies with the {\it Sleeping
Beauty} problem, as discussed for example by Adam Elga in
Ref.~\cite{elga_00}. In that scenario the number of awakenings of
Sleeping Beauty is determined by the flip of a coin, while in
this problem the number of observer-instants with a given
subjective state is determined by the random choice of which
universe $U_\alpha$ is realized.  The conclusion that we describe
here is analogous to the ``thirder'' position of the {\it
Sleeping Beauty} problem (i.e., giving 1/3 as Beauty's credence
that the coin in the thought experiment landed heads, in a
scenario where heads leads to one awakening (on Monday), and
tails leads to two (on Monday and Tuesday).) The observer-instant
numbers in our scenario are analogous to the days of the week in
{\it Sleeping Beauty.} The ``thirder'' answer is in accordance
with the understanding that if one outcome leads to more
observations than another, then the probability of observing that
outcome is enhanced. The logic that we used here to justify our
conclusion is in fact a generalization of the argument given by
Elga in Ref.~\cite{elga_00}.  Physicists whom we know generally
take the ``thirder'' position, as exemplified by Joe Polchinski's
blog post \cite{carroll_blog}.}

The ``weighting by the number of observers'' is incorporated in
the factor $\nprop(U_\alpha,\calS,\calP)$ in the numerator of
Eq.~(\ref{EQ:P1pP2}), which can be seen to arise in
Eq.~(\ref{EQ:nprop}) from the fact that each observer-instant
$\,\calI_i$ contributes to the overall probability.

\section{Comparing prescriptions in simple box models}
\label{APP:BoxMod}

In this Appendix we consider a simple ``box model'' universe of
the same general type that was considered by Hartle and
collaborators in Refs.~\cite{hartle+srednicki_07},
\cite{srednicki+hartle_10}, \cite{hartle+hertog_15}, and
\cite{hartle+hertog_16}.  
(In Refs.~\cite{hartle+srednicki_07} and
\cite{srednicki+hartle_10}, the boxes were referred to as
``cycles in time''.) These calculations will serve as a concrete
example of how such models are treated by the {\it Method of Many
Repetitions} that we advocate, and they will also provide an
opportunity to compare with several different methods discussed
by Hartle and collaborators in the papers mentioned above.

\begin{figure}
   \includegraphics[width=3.5in]{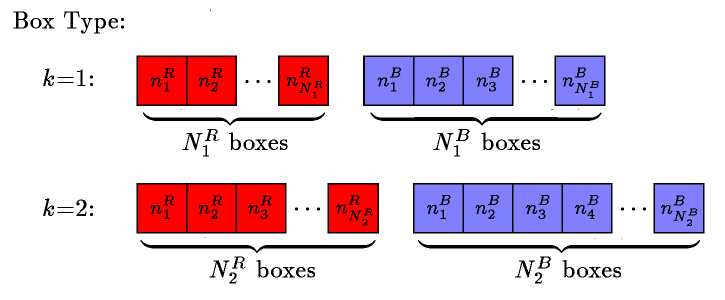}
	\caption{A diagram to illustrate the box model theories
     described in
     Appendix~\ref{APP:BoxMod}.}
	\label{FIG:box}
\end{figure}

The box model that we will consider, illustrated in
Fig.~\ref{FIG:box}, is a generalization of the box models
considered in the Hartle et al.\ papers mentioned above.  The
model is described by the following assumptions:

\begin{itemize}

\item[(a)] The theories $\calT$ that we will consider will have
the form described by Fig~\ref{FIG:box}.  Each theory will allow
model universes with either of two {\it box types}, labeled by
$k=1$ or $k=2$.  Each $k=1$ universe will consist of $N^R_1$ red
boxes and $N^B_1$ blue boxes.  These boxes might be thought of as
being separated from each other in time or in space (or
both)---it doesn't matter.  Similarly, each $k=2$ universe will
consist of $N^R_2$ red boxes and $N^B_2$ blue boxes.

\item[(b)] Each box of each model universe might or might not
contain observers.  (Time dependence is neglected in these simple
box models, so we make no distinction between observers and
observer-instants.) For simplicity, we will assume that all the
observers are identical, and in particular are in the same
subjective state $\calS$.  We will also assume that they can
measure the color of their box without error.  The number of
observers in the $i$'th red box will be denoted by $n^R_i$, where
$i$ ranges from 1 to the number of red boxes, $N^R_k$. 
Similarly, the number of observers in the $j$'th blue box will be
denoted by $n^B_j$, where $j$ ranges from 1 to the number of blue
boxes, $N^B_k$.  To make contact with the notation that we used
in Sec.~\ref{SEC:PIMMR2}, a complete description of a model
universe $U_\alpha$ is obtained by equating $\alpha$ with the
multiplet of parameters specified above:
\begin{equation}
\alpha \equiv (k, n^R_1, \ldots , n^R_{N^R_k}, n^B_1, \ldots,
     n^B_{N^B_k}) \ .
\label{EQ:alpha}
\end{equation}

\item[(c)] Each theory $\calT$ specifies the numbers 
$N^R_1$, $N^B_1$, $N^R_2$, $N^B_2$ of each type and color of box,
and also specifies the probabilities $\Pcos(U_\alpha)$ for any
possible universe $U_\alpha$.  The probability that the box type
is $k=1$ will be denoted by $p_1$, while the probability for
$k=2$ will be $p_2 = 1-p_1$.  We will assume that the probability
that any red box contains exactly $n$ observers is given by the
function $p^R_\obs(n)$, and similarly the probability that any
blue box contains exactly $n$ observers is given by
$p^B_\obs(n)$.  Thus,
\begin{equation}
\Pcos(U_\alpha) = p_k \prod_{\rm red\ boxes}
   p^R_\obs(n_i{^R})
   \prod_{\rm blue\ boxes} p^B_\obs(n_j{^B}) \ .
   \label{EQ:probUbox}  
\end{equation}
The quantities $N^R_1$, $N^B_1$, $N^R_2$, $N^B_2$, $p_1$, and
$p_2$, and the functions $p^R_\obs(n)$ and $p^B_\obs(n)$ all
depend on the theory $\calT$ under consideration. 

\end{itemize}

We now imagine that I am one of the observers, and that I measure
the color of my box and find that it is red.  To update the
probabilities that I assign to various theories, I need to
calculate the first-person probability that I should see red.

Using the {\it Method of Many Repetitions}, I would start by
calculating the expectation value for the number of observers in
the ensemble who will see red.  If the ensemble contains $N$
model universes, this expectation value is given by
\begin{equation}
   \expect{n^R_\ens}
     = N \left[ p_1 N^R_1 + p_2
     N^R_2 \right] \bar n_{\rm red} \ ,
\end{equation}
where $\bar n_{\rm red}$ is the expected number of observers in
any single red box, as calculated from $p^R_\obs(n)$.  That is,
\begin{equation}
   \bar n_{\rm red} = \sum_{n=0}^\infty n \, p^R_\obs(n) \ .
   \label{EQ:nbar}
\end{equation}
There will be an analogous expression for $\expect{n^B_\ens}$,
and the expectation value for the total number of observers is
just the sum of the two expressions.  For the asymptotic
ensemble, which corresponds to the limit as $N \rightarrow
\infty$, the ratios of the actual numbers of observers who see
red or blue to these expectation values should approach 1 (law of
large numbers).  The first-person probability that I should see
red is equal to the fraction of observers in the asymptotic
ensemble who see red, which is then given by
\begin{equation}
   \pfirst({\rm red}|\calT) = \frac{1}{Z} \left[ p_1
     N^R_1 + p_2 N^R_2 \right] \bar n_{\rm red} \ ,
\label{EQ:pfirstbox}
\end{equation}
where
\begin{equation}
   \begin{aligned}
   Z &\equiv  \left[ p_1
     N^R_1 + p_2 N^R_2 \right] \bar n_{\rm red} \\
   & \qquad + \left[ p_1
     N^B_1 + p_2 N^B_2 \right] \bar n_{\rm blue}\ .
   \end{aligned}
\end{equation}

This answer can be contrasted with several different approaches
taken by Hartle and collaborators.  In
Ref.~\cite{hartle+srednicki_07} (hereafter HS2007), Hartle and
Srednicki introduced a box model which can be related to our
model by taking $p_1=1$, $p_2=0$, so there is only one box type. 
The probability that any box contains one or more observers is
denoted by $p_E$, so $p_E = 1 - p^R_\obs(0) = 1 - p^B_\obs(0)$. 
The authors consider two theories for this universe: one called
{\it all red} ($AR$), for which $N^B_1(AR)=0$, so all the boxes
are red, and one called {\it some red} ($SR$), where both
$N^R_1(SR)$ and $N^B_1(SR)$ are nonzero.  For this special case,
the answer we gave in Eq.~(\ref{EQ:pfirstbox}) reduces to
\begin{equation}
\pfirst({\rm red}|\calT) = \frac{N^R_1 \bar n_{\rm red}}{N^R_1
\bar n_{\rm red} + N^B_1 \bar n_{\rm blue}}\ .
\end{equation}
In analyzing this model, HS2007 adheres to what we have dubbed
the {\it Principle of Required Certainty}, as we discussed in
Sec.~\ref{SEC:PRC}\null.  In this approach first-person
probabilities are irrelevant\null.  If an observer measures the
color of his box and finds it to be red, he updates his belief in
the relevant theories using only the third-person statement that
there exists, somewhere in the universe, at least one red ($R$)
box in which observers exist ($E$).  This probability is given by
\begin{equation}
P(E,R | \calT) = 1 - (1-p_E)^{N^R_1(\calT)} \ .
\end{equation}
If $p_E > 0$ and $N^R_1(\calT)$ is very large, this probability
approaches 1 in both the $AR$ and $SR$ theories, even if the
boxes in the $SR$ theory are almost all blue.  By contrast,
updates based on our Eq.~(\ref{EQ:pfirstbox}) would under these
circumstances strongly favor $AR$.  HS2007 rejects our approach,
since in their view there is no reason to believe that the color
that we see is in any sense representative:
\begin{quote}
In such cases it is tempting to reason as follows: In the $SR$
theory with mostly blue cycles we are more likely to observe blue
than red. Since we observe red we should reject this theory.
\dots This, however, is an instance of the selection fallacy, and
in more familiar situations leads to absurd conclusions
\dots~.
\end{quote}
We responded to this {\it selection fallacy} argument in
Sec.~\ref{SEC:Fallacy}.

Srednicki and Hartle \cite{srednicki+hartle_10} (hereafter
SH2010) returned to this model a few years later, at this point
recognizing that the method proposed in HR2007 is ``not effective
for discriminating between theories in a very large universe,''
and that therefore alternatives should be sought.  So they
proposed using first-person probabilities, with an assumed
xerographic distribution.  Although they consider it to be only
one out of many possible choices, they do consider the
xerographic distribution in which we are typical of all the
observing systems that exist, as would be implied by the
PSLI\null.  Nonetheless their answer does not agree with ours,
because they treat the possible nonexistence of observers in a
way that makes no sense to us.  We all agree that first-person
probabilities are the probabilities for ``what {\it we} will
measure,'' (SH2010, p.~1), but they nonetheless attribute a
nonzero probability that our measurements will show that no
observers exist.  Their Eq.~(5.4),
\begin{equation}
P^{(1p)}(E|\calT,\xi^{\rm typD}) = 1 - (1-p_E)^{N_R(\calT)} \ ,
\end{equation}
gives a probability less than one to the proposition that at
least one observer exists.  The first-person likelihood of seeing
red, to be used in Bayesian updates, is given by
\begin{equation}
P^{(1p)} (E,R | \xi^{\rm typO},\calT) =
     \frac{N_R(\calT)}{N_R(\calT)+N_B(\calT)} \left[ 1 - (1 -
     p_E)^N \right] \ .
\end{equation}
This agrees with our Eq.~(\ref{EQ:pfirstbox}) except for the
factor in brackets, and the factors of $\bar n_{\rm red}$ and
$\bar n_{\rm blue}$ that appear in our answer, due to our
weighting by observers.  According to the answer in SH2010, the
three first-person probabilities---that we see red, that we see
blue, and that we don't exist---sum to one.  The comparison of
the answers also shows the impact of our weighting by observers. 
If $N^R_1 = N^B_1$, but every red box contains a million
observers and every blue box contains only one observer, then our
method will strongly favor our seeing red, while the method of
HS2010 will assign equal probabilities.

In Ref.~\cite{hartle+hertog_15} (hereafter HH2015), written
somewhat later, Hartle and Hertog introduce a slightly more
complicated box model.  (A similar model is discussed in
Ref.~\cite{hartle+hertog_16}.) In this model, $p_1$ and $p_2$ are
both nonzero, with $N^B_1=0$ and $N^R_2=0$.  Assuming a uniform
xerographic distribution, HH2015 found that the first-person
probability for seeing red is given by
\begin{equation}
\pfirst({\rm red}|\calT) = \frac{p_1 \left[1 - (1-p_E)^{N^R_1}\right]}
     {p_1 \left[1 - (1-p_E)^{N^R_1}\right] +
     p_2 \left[1 - (1-p_E)^{N^B_2}\right]}\ , 
\end{equation}
with a similar expression for $\pfirst({\rm blue}|\calT)$. These
two probabilities sum to one, as they should, with no allowance
for a first-person probability of not existing. This answer again
does not agree with our Eq.~(\ref{EQ:pfirstbox}), but the
difference can be traced to a single difference in our method of
calculation.  That is, the {\it Method of Many Repetitions}
assigns a weight to each possible universe $U_\alpha$
proportional to its third-person probability and to the number of
observers, while the HH2015 method weights each possible universe
only by its third-person probability, restricted to universes
that have at least one observer.

\bibliography{ProbInf-arXiv}
\end{document}